\documentclass[aps,prl,twocolumn,floatfix,amsmath,amssymb,superscriptaddress]{revtex4}
\usepackage{epsfig}
\usepackage{graphicx}
\usepackage{dcolumn}
\usepackage{bm}
\usepackage{color}
\usepackage[paperwidth=210mm,paperheight=297mm,centering,hmargin=2.2cm,vmargin=2.2cm]{geometry}

\begin{document}

\title{Enhancement of long range correlations in a 2D vortex lattice by incommensurate 1D disorder potential}

\author{I. Guillam\'on}
\affiliation{Laboratorio de Bajas Temperaturas, Departamento de F\'isica de la Materia Condensada, Instituto de Ciencia de Materiales Nicol\'as Cabrera, Condensed Matter Physics Center, Universidad Aut\'onoma de Madrid, E-28049 Madrid, Spain}
\affiliation{Unidad Asociada de Bajas Temperaturas y Altos Campos Magn\'eticos, UAM, CSIC, Cantoblanco, E-28049 Madrid, Spain}
\affiliation{H.H. Wills Physics Laboratory, University of Bristol, Tyndall Avenue, Bristol, BS8 1TL, U.K.}
\author{R. C\'ordoba}
\affiliation{Laboratorio de Microscop\'ias Avanzadas (LMA), Instituto de Nanociencia de Arag\'on (INA), Universidad de Zaragoza, E-50018 Zaragoza, Spain}
\affiliation{Departamento de F\'isica de la Materia Condensada, Universidad de Zaragoza, 50009 Zaragoza, Spain}
\author{J. Ses\'e}
\affiliation{Laboratorio de Microscop\'ias Avanzadas (LMA), Instituto de Nanociencia de Arag\'on (INA), Universidad de Zaragoza, E-50018 Zaragoza, Spain}
\affiliation{Departamento de F\'isica de la Materia Condensada, Universidad de Zaragoza, 50009 Zaragoza, Spain}
\author{J.M. De Teresa}
\affiliation{Laboratorio de Microscop\'ias Avanzadas (LMA), Instituto de Nanociencia de Arag\'on (INA), Universidad de Zaragoza, E-50018 Zaragoza, Spain}
\affiliation{Departamento de F\'isica de la Materia Condensada, Universidad de Zaragoza, 50009 Zaragoza, Spain}
\affiliation{Instituto de Ciencia de Materiales de Arag\'on (ICMA), Universidad de Zaragoza-CSIC, Facultad de Ciencias, Zaragoza, 50009, Spain}
\author{M.R. Ibarra}
\affiliation{Laboratorio de Microscop\'ias Avanzadas (LMA), Instituto de Nanociencia de Arag\'on (INA), Universidad de Zaragoza, E-50018 Zaragoza, Spain}
\affiliation{Departamento de F\'isica de la Materia Condensada, Universidad de Zaragoza, 50009 Zaragoza, Spain}
\author{S. Vieira}
\affiliation{Laboratorio de Bajas Temperaturas, Departamento de F\'isica de la Materia Condensada, Instituto de Ciencia de Materiales Nicol\'as Cabrera, Condensed Matter Physics Center, Universidad Aut\'onoma de Madrid, E-28049 Madrid, Spain}
\affiliation{Unidad Asociada de Bajas Temperaturas y Altos Campos Magn\'eticos, UAM, CSIC, Cantoblanco, E-28049 Madrid, Spain}
\author{H. Suderow}
\affiliation{Laboratorio de Bajas Temperaturas, Departamento de F\'isica de la Materia Condensada, Instituto de Ciencia de Materiales Nicol\'as Cabrera, Condensed Matter Physics Center, Universidad Aut\'onoma de Madrid, E-28049 Madrid, Spain}
\affiliation{Unidad Asociada de Bajas Temperaturas y Altos Campos Magn\'eticos, UAM, CSIC, Cantoblanco, E-28049 Madrid, Spain}
\begin{abstract}
\textbf{Long range correlations in two-dimensional (2D) systems are significantly altered by disorder potentials. Theory has predicted the existence of disorder induced phenomena such as Anderson localization\cite{Anderson58} and the emergence of novel glass and insulating phases from the competition between interactions and disorder as for example the Bose glass\cite{Fisher89}. More recently, it has been shown that disorder breaking the 2D continuous symmetry, such as a one dimensional (1D) modulation, can enhance long range correlations \cite{Wehr06}. Experimentally, it remains difficult to find well-controlled model systems. Developments in quantum gases have allowed the study of the interplay between interaction and disorder and the observation of a wealth of phenomena including the transition between superfluid and insulating glassy states\cite{Billy08,Sanchez-Palencia10}. However, there are no experiments exploring the effect of symmetry-breaking disorder. In 2D superconducting vortex lattices, vortex density and interactions can be varied in presence of quenched disorder by changing the magnetic field. Here, we create a 2D vortex lattice at 0.1 K in a superconducting amorphous thin film with a well defined 1D thickness modulation and track the field induced modification in the vortex arrangements using scanning tunneling microscopy. We find that the 1D modulation becomes incommensurate to the vortex lattice and drives an order-disorder transition, behaving as a scale-invariant disorder potential. Through direct visualization of individual vortices, we show that the transition occurs in two steps and is mediated by the proliferation of topological defects. We calculate orientational and positional correlation functions and critical exponents. We find that they are far above theoretical expectations for scale-invariant disorder \cite{Nattermann95,Fertig95,HN79} and that, unexpectedly, they follow instead the critical behaviour which describes dislocation unbinding melting\cite{Berezinskii72,kosterlitz73}. Our data show for the first time that randomness disorders a 2D crystal, and evidence the transformation induced by symmetry breaking disorder in interactions and the critical behaviour of the transition.}
\end{abstract}

\maketitle

The competition between order and disorder is a fundamental problem in condensed matter physics. New insights impact directly the understanding of disorder induced phenomena in many different systems such as crystalline solids\cite{Fertig95,Carpentier98}, electronic or magnetic arrangements\cite{Natterman90}, localization in metals and superconductors or vortex lattices in superconductors and condensates \cite{Minnhagen87,Billy08}. In 2D, long wavelength disorder induces deviations in the atomic positions from the perfect lattice with the mean squared displacement diverging logarithmically at large distances\cite{Mermin68}. One major consequence is the so-called Mermin-Wagner-Hohenberg (MWH) theorem\cite{Mermin68,Hohenberg67b} which states that no true order exists in 2D at any finite temperature. Usually, we can distinguish between static quenched disorder and fluctuations. In absence of quenched disorder, thermal fluctuations drive the 2D melting transition which is described by BKTHNY theory through the two-stage proliferation and unbinding of topological defects\cite{Berezinskii72,kosterlitz73,HN78,Young79}. Quenched disorder, on the other hand, is expected to suppress long range correlations more effectively than temperature\cite{Nelson83}. It can be classified as pinning with identifiable length scales, such as impurities or defects in 2D crystals or as scale-invariant (random) disorder as for example in an amorphous film. Pinning destroys long range 2D correlations at any strength\cite{Giamarchi00,Guillamon11a}. Scale-invariant disorder produces power law decaying correlations and a transition to a disordered lattice with exponentially falling correlations above a critical disorder strength\cite{Nattermann95,Fertig95}. The order-disorder transition induced by scale invariant disorder has been investigated in a wide range of physical systems such as 2D disordered XY models\cite{Nattermann95}, 2D solids\cite{Fertig95}, Josephson junction arrays\cite{Nattermann96}, colloids or Lennard-Jones systems\cite{SadrLahijany97}. However, the disorder mechanism ---the way disorder proliferates at zero temperature--- has not been observed directly. Disorder induced order has been recently proposed when quenched disorder breaks the continuous 2D symmetry. This occurs, for example, by introducing a 1D periodic disorder potential\cite{Wehr06}. Within this scenario, true long range order may be favored by the 1D disorder, breaking MWH theorem. Calculations show the stabilization of the quantum Hall ferromagnetic state in graphene monolayers due to strain-induced easy-plane anisotropy\cite{Abanin07} or improved control of the relative phase in randomly coupled condensates\cite{Niederberger08}. The experimental realization of such a disorder-induced order in absence of thermal fluctuations has not been reported yet. The effect of symmetry breaking on microscopic properties and the critical exponents of the order-disorder transition are unknown. 

\hspace*{0.4cm}

Here we shed light on these questions by directly imaging the order-disorder transition in a 2D vortex lattice induced by a 1D periodic potential. By changing the magnetic field, we modify the coupling strength between the 1D periodic potential and the vortex lattice as well as the intervortex distance $a_{0}=(3/4)^{1/4}(\phi_{0}/B)^{1/2}$. This allows us to go from a locked 2D solid where the lattice is commensurate to the 1D potential (Fig. 1a, left) to a floating 2D solid at larger densities where it becomes incommensurate with the 1D modulation. In the latter case, the 1D modulation behaves as quasi-random quenched disorder for the vortex lattice and drives an order-disorder transition (Fig.1a, right).

\hspace*{0.4cm}
\begin{figure*}
\includegraphics[width=16cm,clip]{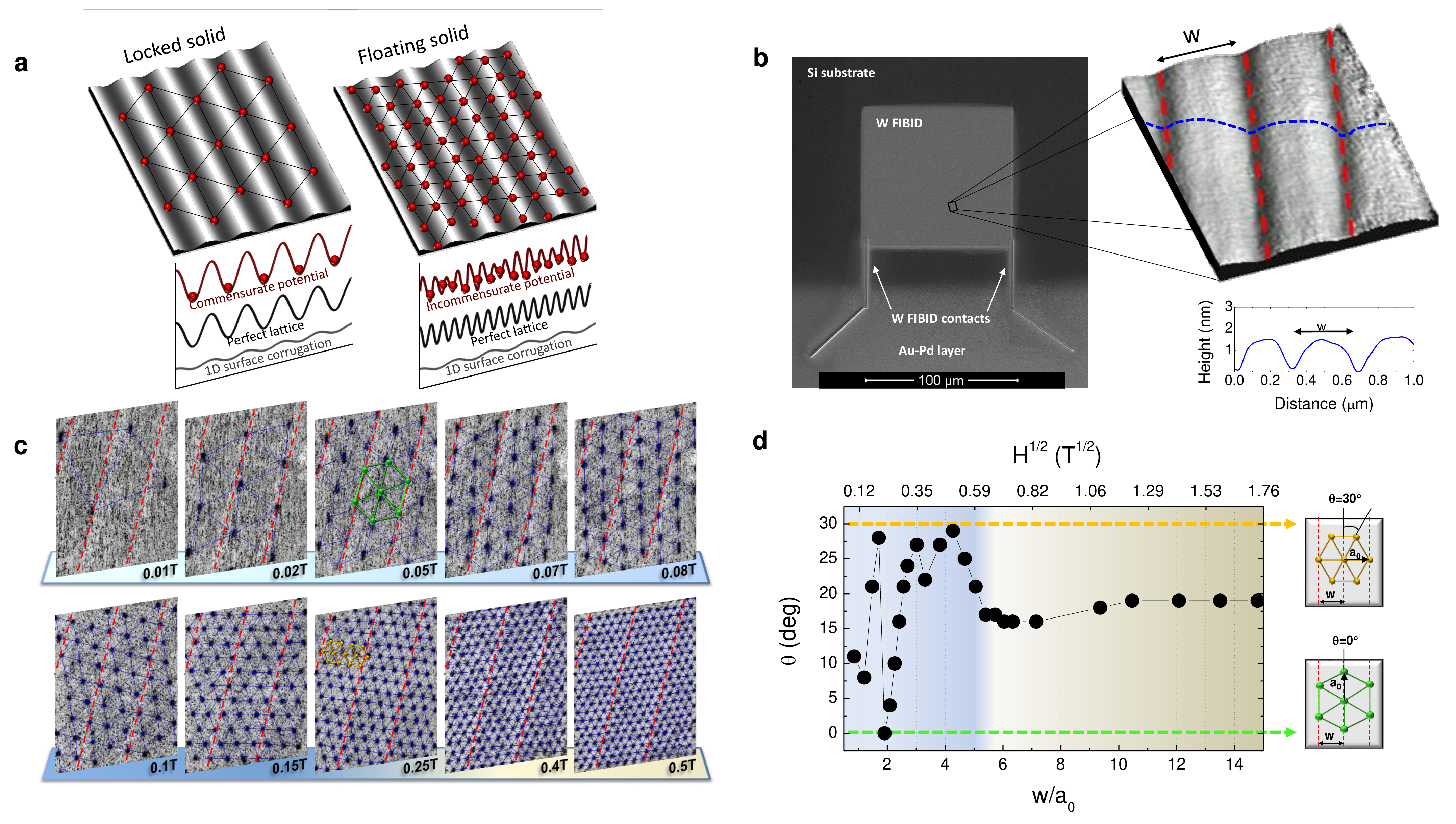}
\caption{\small{\textbf{Unlocking of the 2D vortex lattice on a linear potential.} \textbf{a} When lattice constant disorder wavelength are similar (small p), commensurate lattices with one crystal axis along the 1D modulation, and incommensurate lattices oriented at an arbitrary angle may appear. Generally, the commensurate lattices are favored, because they lower the elastic energy of the lattice. Commensurate arrangements are locked to the 1D potential (as shown in the left panel). When the lattice constant is much smaller than the 1D modulation, the gain in elastic energy obtained by adjusting to the potential decreases and the lattice can show incommensurate configurations. The lattice is unlocked and produces a floating solid (as shown in the right panel). \textbf{b} Scanning Electron Microscopy image of the sample (left panel). Sample is biased through the contact pads shown in the bottom of the panel (see SI). Top right panel shows a STM topography of a $1\times 1.2$ $\mu m^{2}$ area, and bottom right panel a cut through it. Red dotted lines mark the 1D modulation. \textbf{c} Sequence of vortex lattice images taken at 0.1 K when increasing the magnetic field up to 0.5 T (see Supplementary video S1 for the whole sequence). Red dashed lines are again the 1D modulation and vortices are shown as black points. Blue lines are Delaunay triangulation. \textbf{d} Dependence of the angle between the 1D modulation and the vortex lattice with the magnetic field. Lines joining points are a guide to the eye. When $p<5$ ($H < 0.4$ T; blue background), the vortex lattice oscillates between the two primary commensurate arrangements $\theta=0^{\circ}$ (green dotted line) and $\theta=30^{\circ}$ (yellow dotted line). This satisfies, respectively, $w=n\sqrt{3}a_{0}/2$ and $w=ma_{0}$, with n and m integers ($n=1$ and $m=1$ are highlighted in upper (yellow) and lower (green) left panels of \textbf{d}). $n=1$ and $m=4$ commensurate arrangements are identified in the vortex images of \textbf{c} at respectively 0.05 T and 0.25 T. At fields above 0.4 T  ($w>p$; khaki coloured in \textbf{c} and \textbf{d}, the lattice unlocks and adopts an orientation independent on 1D potential that does not show angular variations with field any more.}}
\end{figure*}

We follow the order-disorder transition by imaging up to thousands of individual vortices from 0.01 T up to close below H$_{c2}$, with the vortex density increasing by a factor of 500. We determine the modifications in the spatial correlation induced by disorder and visualize the microscopic details of ordered and disordered phases. Our experiments show enhanced long range correlations in presence of a 1D modulation. 

\hspace*{0.4cm}

We use Scanning Tunneling Microscopy/Spectroscopy (STM/S) and work at temperatures low enough (0.1 K) to neglect any temperature induced effect. Our sample is an ultra flat amorphous thin film with a thickness, $d$, of 200 nm fabricated by focused-ion-beam-induced deposition (microscopy and nanofabrication methods are described in detail in sections I and II of the Supplementary Information, SI). $d$ is far below than the characteristic length for the vortex bending along the field direction, L$_c$, so that the vortex lattice forms a 2D solid. The surface roughness is below 1\% of $d$ and consists of a smooth 1D modulation with period, $w = 400 nm$ (Fig. 1b). 

\hspace*{0.4cm}

The coupling strength between the vortex lattice and the 1D modulation depends on the commensurability ratio $p$, defined as $p = w/a_{0}$, and the relative orientation between them $\theta$\cite{Radzihovsky01}. Commensurate vortex configurations are expected at $p = n $ or $n\sqrt(3)/2$, with $n$ an integer, and $\theta = 30$ or $0$ degrees, respectively. Fig.1c shows the sequence of vortex lattice images obtained at lower magnetic fields. Below 0.4 T, with $p \lesssim 5$, the lattice suffers a series of commensurate to incommensurate transitions which produces the rotation of its overall orientation between $\theta = 0^{\circ}$ and $30^{\circ}$, while keeping a perfect hexagonal arrangement. Fig. 1d shows $\theta$ as a function of $p$ for the vortex lattice images taken as increasing the magnetic field. As increasing $p$ above 0.4 T ($p \gtrsim 5$), the rotation of the vortex lattice ceases and its orientation becomes independent on the 1D potential. The lattice is no longer commensurate to the 1D potential and forms a floating 2D solid.  

\hspace*{0.4cm}

In Fig. 2a we show the sequence of vortex lattice images between 0.5 T and 5.5 T. We identify three different phases. In phase I, between 0.5 T and 2 T, there are no topological defects, and all vortices are surrounded by six nearest neighbors. However, the vortex positions show small deviations from those expected for a perfect hexagonal lattice which become gradually more pronounced when increasing the magnetic field. Between 2.5 T and 4.5 T, in phase II, we observe the appearance of dislocations, i.e. pairs of 5-fold and 7-fold coordinated vortices. We identify bound dislocation pairs as well as isolated dislocations. Above 4.5 T, in phase III, the density of dislocations experiences a strong increase and we identify the appearance of free disclinations in form of isolated 5-fold or 7-fold coordinated vortices. The images at 5 T and 5.5 T show that defects are over the whole sample, and produce a disordered vortex lattice. 

\hspace*{0.4cm}

One important observation ---the appearance of fluctuations in the local vortex density $\rho$--- is shown in Fig.2b-c. The standard deviation in $\rho$ grows with the field-induced proliferation of defects from less than 5\% in phase I to up to 20\% at 5.5 T in phase III (Fig. 2c). Density fluctuations are characteristic of long wavelength or fully uncorrelated quenched disorder potential\cite{Giamarchi95}.

\begin{figure*}
\includegraphics[width=16cm,clip]{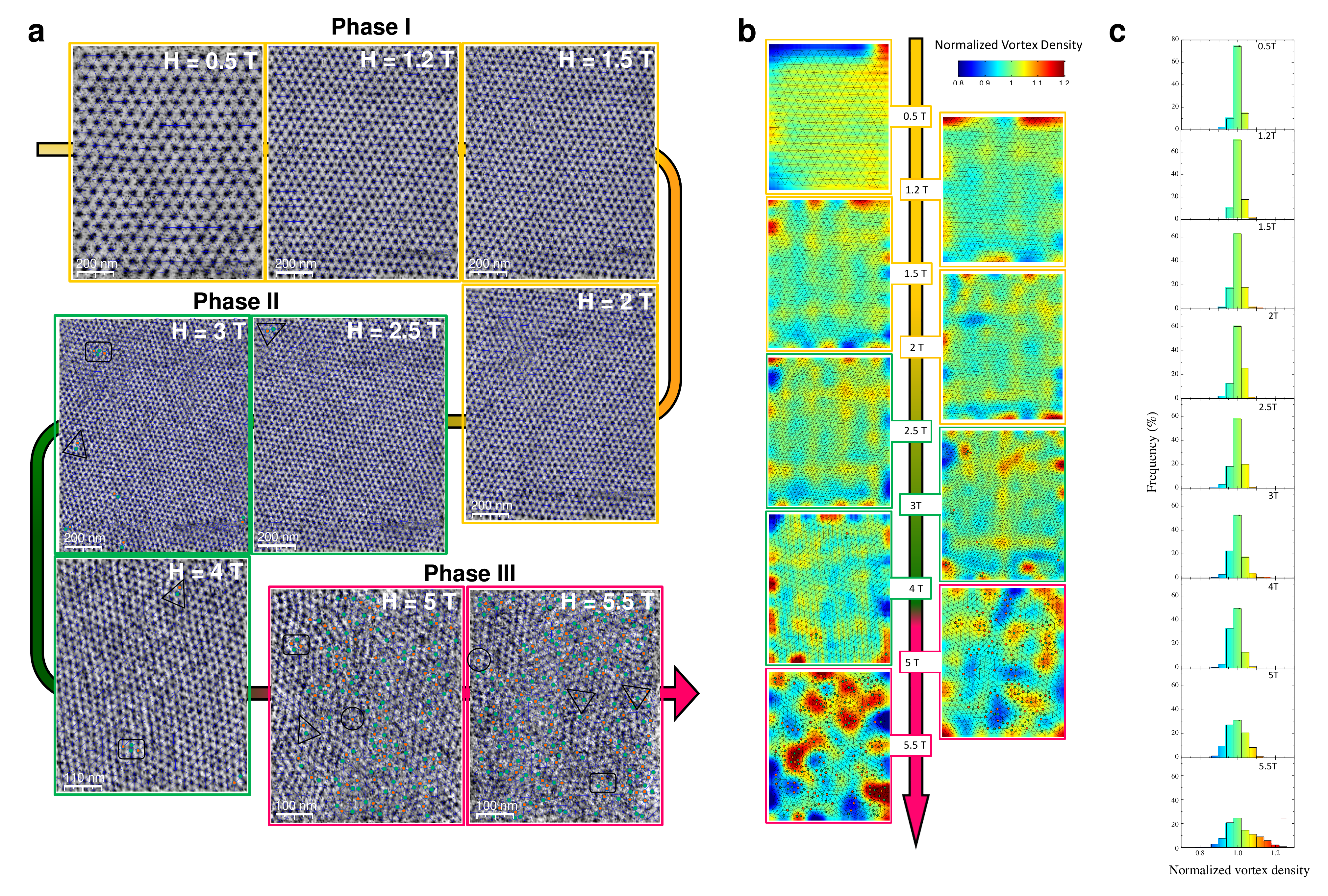}
\caption{\small{\textbf{Order-disorder transition in the 2D-vortex lattice at 0.1 K.} \textbf{a} Vortex lattice images taken at 0.1 K when increasing the magnetic field from 0.5 T to 5.5 T (see Supplementary video S2). Vortices are shown by black points and the Delaunay triangulated lattice as blue lines. Vortices with five and seven nearest neighbours are identified by green and orange points. Dislocations, formed by five and seven nearest neighbour pairs of vortices are identified as black triangles, pairs of dislocations as black rectangles and isolated disclinations as black circles.  Above 4 T the image size is decreased to better resolve individual vortices. Average intervortex distance $a_{0}$ decreases with field following the expected dependence in a triangular vortex lattice (see SI). Evolution of the disordering process in the reciprocal space is shown in Section IV of the SI through the gradual changes of the height and width of the vortex lattice Bragg peaks. \textbf{b} Positional fluctuations of the vortex density over the disorder process calculated from the vortex images shown in \textbf{a}. We use the color scale given on the top right panel (see section \textit{III.3} in SI). Deviations in the local vortex density become stronger with increasing the magnetic field. Histograms of the vortex density obtained from the images in \textbf{b} are shown in \textbf{c} using the same color scale.}}
\end{figure*}
\hspace*{0.4cm}

To further quantify the spatial dependence of vortex disorder, we calculate translational and orientational correlation functions, $G_{K}(r)$ and $G_{6}(r)$, directly obtained from the individual vortex positions in the images.  Deviations with respect to the perfect lattice produce a decay with $r$ of the envelope of $G_{K} (r)$ and $G_{6} (r)$ that describes, respectively, weakening of translational and orientational correlations. Fig. 3a shows the evolution with the magnetic field of $G_{K} (r)$ and $G_{6}(r)$. In phase I, between 0.5 T and 2 T, we observe that $G_{6} (r)$ remains close to 1 and independent of the distance, whereas $G_{K} (r)$ decays following a power-law dependence, $G_{K}(r)\sim r^{-\eta_{K}}$, with $\eta_{K}$ increasing with field. Above 2 T, in phase II, $G_{K}(r)$ decays exponentially at large distances when $\eta_K=1/3$, and when a finite amount of defects starts to be observed in the images. $G_{6} (r)$ shows a power-law decay, $G_{6}(r)\sim r^{-\eta_{6}}$, with $\eta_{6}$ continuously increasing from 2 T up to 4.5 T. In phase III, above 5 T, $G_{6} (r)$ decays exponentially when $\eta_6=1/4$, and the defect density diverges, reaching 0.4, i.e. nearly half of all vortices have 5 or 7 nearest neighbours at 5.5 T (Fig. 3b).  The observed behaviour follows the microscopic two-step sequence for the proliferation of disorder described by BKTHNY theory, with critical values for the exponents $\eta_{K}^{c}=1/3$ and $\eta_{6}^{c}=1/4$\cite{Zahn99}.

\hspace*{0.4cm}

To investigate the microscopic disorder mechanism, we further analyse the first entrance of disorder in the ordered phase I. Deviation in the vortex positions with respect to perfect hexagonal lattice can be quantified by the relative displacement correlator $B(r)$ ($B(r)=\langle [\textbf{u}(\textbf{r})-\textbf{u}(\textbf{0})]^{2}\rangle/2$ where $\textbf{u}(\textbf{r}) = \textbf{r}-\textbf{r}_{p}$ is the displacement of each vortex at $\textbf{r}$ from its position in the perfect lattice $\textbf{r}_{p}$\cite{Giamarchi95}). We find that $B(r)$ grows as $ln(r/a_{0})$ in the dislocation-free Phase I. In Fig. 3c we plot the result at 1.5 T (see SI for details in calculations and B(r) at different fields in Phase I). In 2D systems, this is the expected behaviour in response to a scale-invariant disorder\cite{Giamarchi00}. Then, we fit the data using the Gaussian approximation $G_{K}(r) = e^{-K^{2}B(r)/2}$ (valid for Gaussian disorder potentials) and translational correlation function $G_{K}(r)$ (shown in Fig. 3a). The comparison reveals a very good agreement (Fig. 3c). Therefore, three independent observations (local vortex density fluctuations, logarithmic growth of $B(r)$ and Gaussian distribution of displacements) show that a random potential drives the transition.

\begin{figure*}
\includegraphics[width=13cm,clip]{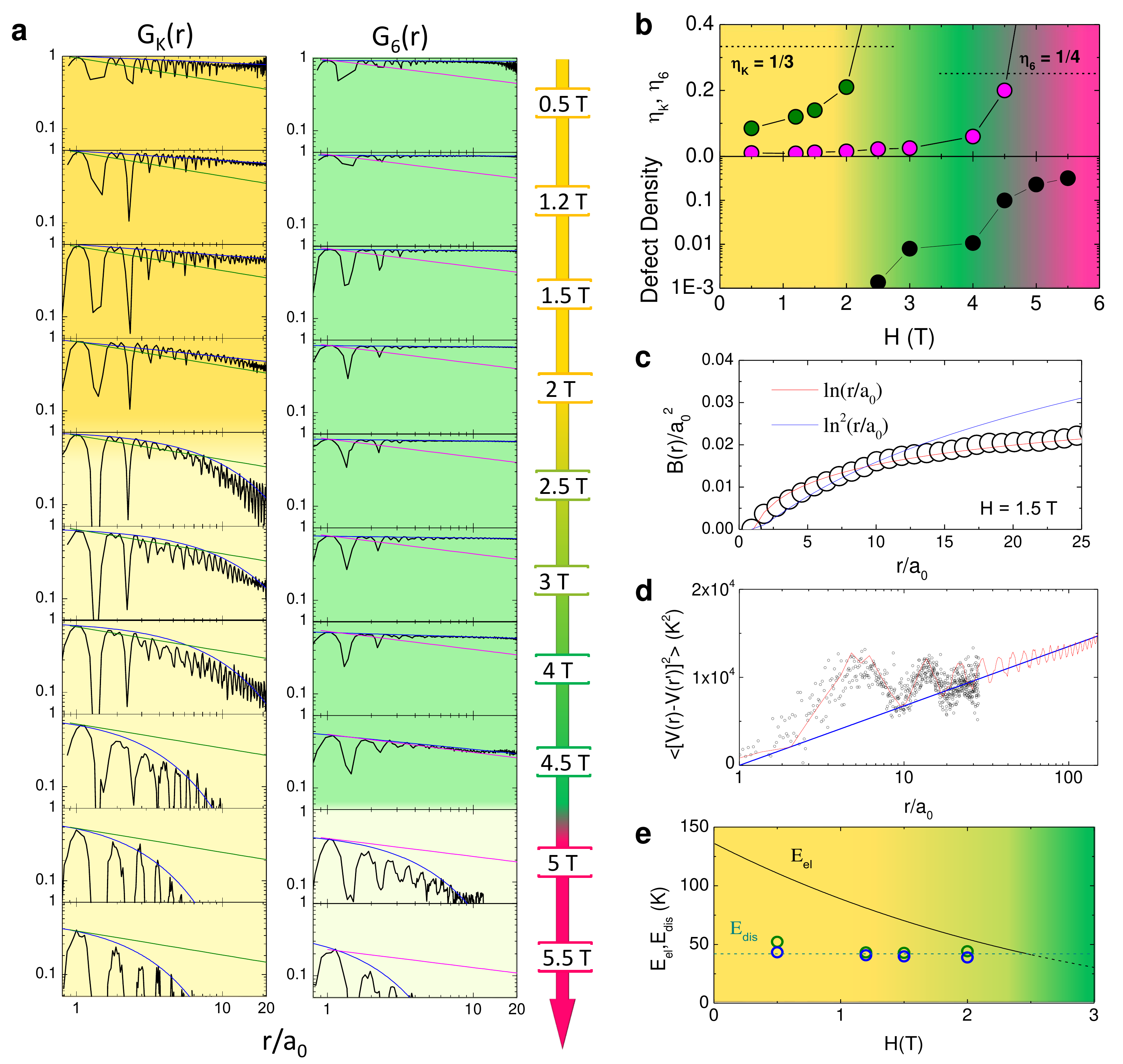}
\caption{\small{\textbf{Correlation functions and critical exponents}. \textbf{a} Field induced changes in the correlation functions for the translational order ($G_{K}(r)$ in left panel) and orientational order ($G_{6}(r)$ in right panel) between 0.5 T and 5.5 T. Blue lines in the graphs are the fits for the envelopes of the correlation functions (black lines). Green and magenta lines in left and right panels stand for the power decay law with, respectively, the critical exponents $\eta_{K}^{c}=1/3$ and $\eta_{6}^{c}=1/4$. The background color (yellow for $G_{K}(r)$ and green for $G_{6}(r)$) represents the range of the order and turns from darker (long-range) to lighter (short-range) when the the exponents $\eta_{K}$ and $\eta_{6}$ become higher than the critical values, and correlation functions change from power-law to exponential decay with the distance $r$. The distance in the X-axis is given in units of $a_{0}$ and peaks in the correlation functions appear at the distances to n-th nearest neighbours. Note that these peaks are well defined when the order is long ranged and become blurred when it turns to be of short range. \textbf{b} The top panel shows the field dependence of the power-law decay exponents $\eta_{K}$ (green points) and $\eta_{6}$ (magenta points). The crossover between phase I (yellow), phase II (green) and phase III (magenta) is determined at the fields where the translational and orientational order become of short range and the exponents $\eta_{K}$ and $\eta_{6}$ reach the critical values (dotted lines in the figure). The lower panel shows the density of five-fold and seven-fold coordinated vortices as a function of the magnetic field. \textbf{c} Correlation function $B(r)/a_{0}^{2}$ at 1.5 T (black dots, see text and SI for details). Red line is $\approx (3\eta_{K}/8\pi^{2}) ln(r/a_{0})$ obtained within the Gaussian approximation for a scale invariant random potential (see text and SI section VI). Blue line gives the $ln^2(r)$ dependence expected within correlated pinning disorder. \textbf{d} Spatial dependence of mean squared correlation of 1D surface potential $\langle[V_{1D}(r)-V_{1D}(r')]^{2}\rangle$ at 1.2 T, where $V_{1D}(r)$ is as defined in the text and  $r, r'$ are the actual vortex positions obtained from the vortex image at 1.2 T (see section III.5 in SI for details in calculations). Red line has been obtained by simulating the topography with a periodic modulation and extrapolating vortex position in an area 100 times larger than in the experiment (see SI). The experiment follows well the simulation. Long range logarithmic correlations of $V_{1D}(r)$ and short range periodic modulation are attenuated at large distances. Blue line is the fit to eq.1 in the text. \textbf{e} Magnetic field dependence of  $E_{el}$ (black line) and $E_{dis}$ (blue and green circles). $E_{dis}$ is estimated independently from power law exponents $\eta_{K}$ in positional correlation (green circles) and long range logarithmic correlations of $V_{1D}(r)$ (blue circles). We find very good agreement between them. Transition from Phase I (yellow) to Phase II (green) occurs when $E_{dis}$ increases above $E_{el}$ and dislocations are first observed in the images.}}
\end{figure*}

\hspace*{0.4cm}

We next focus on the source of scale-invariant disorder driving the transition. No thermally or quantum induced fluctuations are available to effectively disorder the vortex lattice here. At 0.1 K, the transition induced by either thermal or quantum fluctuations is expected to occur at a magnetic field extremely close to $H_{c2}$ (see SI). Our sample is compositionally homogeneous both laterally and across its thickness, and amorphous. Thus, no quenched disorder is expected from compositional or structural changes in our film. Thickness variations, which are given here by the 1D modulation, come out as most likely source for quenched disorder. The fundamental property of a scale-invariant potential $V(r)$ is that is has long range logarithmic correlations\cite{Nattermann95}
\begin{equation}
\langle[V(r)-V(r')]^{2}\rangle = 4\sigma J^{2} ln(r-r')
\end{equation}
where $J = \mu d a_{0}^{2}/2\pi$ is the elastic interaction strength (the magnetic field dependence of the shear modulus $\mu$ is discussed in SI) and $\sigma$ is the disorder strength. In Fig. 3d we calculate the spatial correlations of V(r) (first term in eq.1) by taking $V_{1D}(r)=z(r)\cdot\varepsilon_{L}$ where $z(r)$ is the topography and $\varepsilon_{L}= (\phi_{0}^{2}/4\pi\mu_{0}\lambda^{2})ln(\lambda/\xi)$ is the vortex energy per unit length (see SI for details). We find that $V_{1D}(r)$ has long range logarithmic correlations and short range smooth periodic correlations at integer multiple of $w$ which are strongly damped at large distances. Thus, incommensurate 1D correlation behaves as a quasi-random disorder potential.  

\hspace*{0.4cm}

We can write the free energy, $F$, following available renormalization group (RG) theory for random disorder as\cite{Nattermann95,Fertig95}:
\begin{equation}
F = -2T ln(L) + J ln(L) - J\sqrt{\sigma / \sigma_{c}} ln(L)
\end{equation}
where the thermal $E_{th}$ and elastic energy $E_{el}$ (the first and second term, respectively), and the disorder energy $E_{dis}$ (third term) diverge logarithmically with the system size $L$\cite{Nattermann95,Fertig95}. $\sigma_{c}$ is the critical value for the disorder strength. In the ordered phase at low temperatures, the relative strength between $E_{el}$ and $E_{dis}$ determines the algebraical decay of the translational correlations with the exponent $\eta_{K}$ in a hexagonal solid given by\cite{Nattermann96,Giamarchi00}, 
\begin{equation}
\eta_{K} = \dfrac{2}{3}[\frac{T}{J}+\sigma]
\end{equation}
Following the transition from power law to exponential decay in $G_{K}(r)$ we found $\eta_{K}^{c}=1/3$ (Fig. 3a) which corresponds to $\sigma_{c} = 1/2$. We now calculate $E_{dis}$ produced by $V_{1D}(r)$ (using eq.1 and 2) and, independently, from the power law decaying of positional correlation functions (using eq.2 and 3 and $\eta_{K}$ values from Fig.3b). The result obtained at different magnetic fields is  plotted in Fig. 3e (as blue and green circles, respectively) together with the magnetic field evolution of $E_{el}$ (black line).  Note that $E_{th}$ at 0.1K is three orders of magnitude below $E_{el}$ and $E_{dis}$, so it is negligible here. The agreement between $E_{dis}$ determined from exponents in correlation functions $\eta_{K}$ (green circles) and from logarithmic correlations in 1D disorder potential (blue circles) is nearly perfect. This shows that the random uncorrelated potential generated by the 1D incommensurate surface corrugation drives the observed order-disorder transition in the 2D lattice. Fig. 3e shows that $E_{dis}$ increases above $E_{el}$ at the magnetic field where we start to observe dislocations in the images. 

\hspace*{0.4cm}

Finally, let us discuss on the critical behaviour of the observed transition. The order-disorder transition at zero temperature is expected, on the basis of RG calculations and models for random quenched disorder, at a disorder strength of $\sigma_{c}=1/8$ and a critical exponent for the hexagonal lattice $\eta_{K}^{c}=1/12$\cite{Nelson83,Nattermann95,Fertig95,Carpentier00}. Our experiments reproduce closely the expected features for the zero temperature phases as induced by a random disorder, with, in particular, positional fluctuations which increase as $ln(r)$ below the critical disorder and correlations decaying exponentially for high disorder strength. But here we find critical values, $\sigma_{c}=1/2$ and $\eta_{K}^{c}=1/3$, which are far above the ones proposed in theory. The difference between random field theories and our experiments is the symmetry breaking 1D modulation. It produces the disorder through incommensuration and provides the energy scale for the random field driving the transition (Fig.3e). A recent proposal shows that the XY model with 1D symmetry breaking disorder has an increased order parameter at all temperatures\cite{Wehr06}. An earlier work also points out that correlations in the disorder potential enhance the critical value of $\sigma$\cite{Nattermann96}. This strongly suggests that the 1D modulation, by breaking symmetry, modifies the screening of the interactions among dislocations to enhance the critical point and exponents with respect to random field theories.

\hspace*{0.4cm}

Our experiments show that, in presence of the 1D symmetry breaking disorder, the critical exponents increase up to the values expected by BKTHNY and that the microscopic disordering behaviour follows the sequence defined by the two-step thermal melting transition. Inherent to this is the presence of an intermediate hexatic phase and of bound dislocations in the order phase which are not expected within random field models\cite{Velarde09}. In presence of random disorder, it has been shown that the critical behaviour of the disorder transition at zero temperature is not of BKTHNY type\cite{Carpentier00}. The question is why does our experiment follow BKTHNY? To answer this question, one needs to calculate new critical points of the order-disorder transition at zero temperature by taking into account symmetry breaking correlations within randomness and their influence on the renormalization of the parameters involved in the transition. Our experiments show that the 2D solid tends to flow into the BKTHNY behaviour with a reduction of the effect of disorder. The choice of creating disorder in two steps by unbinding and proliferation of topological defects describes possibly more phenomena than just 2D thermal melting.

\hspace*{0.4cm}

Overall, our data represent the first evidence that incommensurate 1D modulation widens the stability range of the ordered phase in 2D system at zero temperature and raise questions that will motivate a detailed examination of the effect of correlations in the critical behaviour of disordered systems. 2D random environments are usually unavoidable in different fields, such as colloids, optical lattices, quantum condensates, 2D crystals or graphene. The experimental approach presented here reveals an exciting new opportunity to produce coherence in the presence of 1D symmetry-breaking fields, as for example nematicity.

\hspace*{1cm}

\paragraph*{\textbf{Acknoledgements}}
This work was supported by the Spanish MINECO (Consolider Ingenio Molecular Nanoscience CSD2007-00010 program,
FIS2011-23488, MAT2011-25046, ACI-2009-0905 and MAT2011-27553-C02), the Comunidad de
Madrid through program Nanobiomagnet (S2009/MAT-1726) and by the Marie Curie Actions under the project FP7-PEOPLE-2013-CIG-618321 and the contract no. FP7-PEOPLE-2010-IEF-273105. We acknowledge technical support of UAM's workshop SEGAINVEX.  

\paragraph*{\textbf{Author Contributions}}
I.G. made the experiment, analysis and interpretation of data. I.G. wrote the paper, together with H.S. and S.V. Samples were made and characterized by R.C. and J.S. J.M.D.T. and R.I. supervised the sample design and fabrication. All authors discussed the manuscript text and contributed to it.

\hspace*{0.2cm}

\paragraph*{\textbf{Competing Interests}} The authors declare that they have no competing financial interests.

\hspace*{0.2cm}

\paragraph*{\textbf{Correpondence}} Correspondence and requests for materials should be addressed to I.G. ~(email: isabel.guillamon@uam.es).

\newpage
\vspace{10cm}
\begin{widetext}
\renewcommand{\figurename}{\textbf{Figure S}}
\begin{center} \textbf{SUPPLEMENTARY INFORMATION} \\ \vskip 0.5cm
\textbf{Enhancement of long range correlations in a 2D vortex lattice by incommensurate 1D disorder potential}
\end{center}
\setcounter{figure}{0} 

\section{I. Scanning Tunneling Microscopy methods}

Imaging of the vortex lattice has been made in a home-made low temperature STM/S that is thermally anchored to the mixing chamber of a dilution refrigerator and inserted into the bore of 9 T superconducting magnet. The experimental set-up, described in detail in Ref. [\textit{S1}], has a linear mechanical positioning system that allows to change the tip's position on top of the sample a distance up to a few mm with a precision in the scale of a few tens of nm. The fine XYZ scanning is made using a conventional piezotube scanner that allows for scanning windows up to $2 \times 2$ $\mu m^{2}$ at low temperatures. Special effort was made to mechanically decouple the STM head from low frequency vibrations of the building and rotatory pumps used for the cooling of the dilution refrigerator. The tip is placed over the center of the W-nanodeposit at room temperature using an optical microscope and its position remains the same within a few micrometers after it is cooled down to mK temperatures. 

\hspace*{0.4cm}

We have improved the mechanical rigidity of the set-up and have been able to make long term measurements over the same scanning window.  This enabled us to study the field induced modifications of vortex lattice in the same area of the sample during several weeks. Note that, thanks to the improved stability, we have been able to make the STM images of the vortex lattice showing the largest amount of vortices.

\hspace*{0.4cm}

The home-made STM-head and electronics provides a resolution of $15$ $\mu eV$ at 0.1 K, tested in low temperature superconductors such as Al [\textit{S1}]. This experimental set-up has been previously used to  image the vortex lattice in the W-based nanodeposits and a number of other superconducting materials [\textit{S2-S4}]. To image the vortex arrangement at fixed temperature and magnetic field, simultaneous topographic and spectroscopic maps of $256 \times 256$ points are measured by taking a full IV curve on each position during the STM scanning of the tip. STS vortex images are obtained by processing the numerical derivative of the IV curves, $\sigma_{V} = dI/dV$, and building maps of the spatial changes of the zero-bias normalized conductance $\sigma_{0}(x,y)$. The superconducting inter-vortex regions are represented as white areas whereas the normal vortex cores, where $\sigma_{0}(x,y)$ is practically equal to 1, as black regions. The STS vortex images shown in the publication and videos are raw and have not been filtered.   

\hspace*{0.4cm}

\section{II. Extra-flat thin films of W-based superconductor}

Superconducting W-based nanodeposits are grown by sweeping a focused Ga$^{+}$ ion beam (FIB) onto a substrate in presence of the metal-organic precursor W(CO)$_{6}$. The Ga$^{+}$ beam dissociates the molecule of the precursor gas adsorbed on the substrate and gives rise to deposits with lateral size and thickness controlled down to nanometric scale. For the sample studied here, the growth process was optimized to reduce the surface corrugation and produce an extra-flat film with a well-defined one dimensional (1D) modulation created by the sweeping FIB.

\hspace*{0.4cm}

Previous STM studies of W-based superconducting thin film were performed in samples with larger surface corrugation [\textit{S2},\textit{S3}]. The films were grown on top of a Au-Pd thin layer used to bias the sample and deposited on a Si substrate. The Au-Pd layer produced a surface corrugation $\Delta d$ of up to 10\% of the total thickness $d$, i.e. of 20 nm. 

\hspace*{0.4cm}

In order to optimize the deposition method, we use a Si/SiO$_{2}$ substrate and control the beam intensity to get a random surface corrugation below an \AA. We have also worked on the pitch of the ion beam to produce the smallest possible and best controlled corrugation due to the sweeping beam. We obtain a perfectly 1D corrugation with a near-sinusoidal form having a period $w$ of 400 nm and a height $\Delta d$ of 2 nm. Remaining corrugation is of atomic size. As the Si/SiO$_{2}$ substrate is insulating, we contact the film by depositing two lines that join the film to an Au-Pd layer deposited previously by thermal evaporation, as shown in the Fig. 1b of the manuscript.

\hspace*{0.4cm}

Homogeneity and composition of the extra-flat nanostructured thin film have been carefully characterized. High resolution transmission electron microscopy, X-ray microanalysis, and X-ray photoelectron spectroscopy show that the deposits are amorphous and that the composition is homogeneous as a function of the depth. The superconducting density perfectly follows simple s-wave BCS behaviour over the whole surface [\textit{S2}]. Fig. \textit{S1} shows a representative curve for the tunneling conductance obtained at 0.1 K and zero magnetic field. Superconducting features are highly homogeneous as a function of the position and give spatial changes below  10\% of the value of the conductance at the quasiparticle peaks.  

\hspace*{0.4cm}
\begin{figure}
\includegraphics[width=10cm,clip]{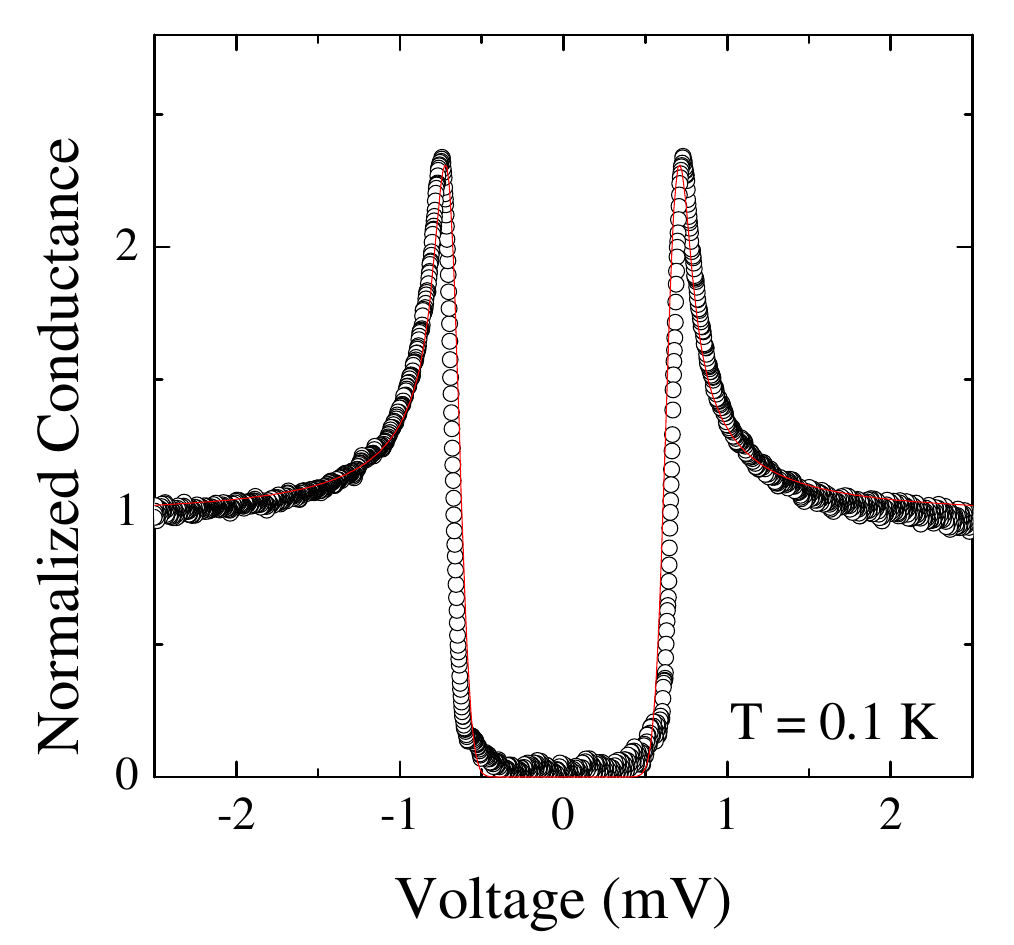}
\caption{\textbf{Superconducting density of states of the extra-flat nanostructured W-film.} Normalized tunneling conductance taken at 0.1 K and 0 T. Red line is the the fit using BCS theory for $\Delta = 0.66 meV$.}
\end{figure} 

\subsubsection*{\textbf{2D character of vortex lattice}}
In the W-nanodeposits, the length scale for the vortex bending parallel to the field direction, $L_{c}$, is of the order of $5$ $\mu m$, which is far above the sample thickness $d = 200$ $nm$ [\textit{S3}]. Thus, the vortices do not bend and make up a perfectly two-dimensional lattice. Actually, this occurs in most thin films of extreme type II superconductors, where the magnetic penetration depth and $L_{c}$ are most often far above the film thickness (see for example [\textit{S5},\textit{S6}]).

\section{III. Methods for numerical analysis of STS vortex images}

\subsection{III.1.Vortex positions and Delanaunay triangulation}
The first step to analyse the vortex lattice images is to determine the coordinates of each vortex in the STS images. To locate the vortex positions, the cores are first identified as regions where the normalized conductance  $\sigma_{0}(x,y)$ is above 0.9 (see section \textit{I}). Then, an algorithm is used to find the centroids of the all the vortices. At the highest fields, where the contrast in the images given by the change of  $\sigma_{0}(x,y)$ between inside and outside the cores is strongly reduced, we first make a low-pass filtering of the vortex images. The result is always thoroughly checked and manually corrected when needed. We estimate that we only obtain a quantifiable error at the highest magnetic fields, where contrast is strongly reduced. There, we may miss or obtain doubtful positions in less than 1\% of the total amount of vortices. For example, in image at 5.5 T (Fig. 2 in the paper), we count 873 vortices and we estimate that our error is below 8 vortices. 

\hspace*{0.4cm}

Delaunay triangulation of vortex positions is calculated using the Delaunay function available in the Matlab library. Usually, at the edges of the images, Delaunay triangulation of the vortex arrangements gives triangles whose vertices do not correspond to vortices being first neighbours. In the Delaunay triangulations presented in the publication, these triangles have been removed from the images. This is relevant because, close to the border, artificial topological defects appear due to the lacking neighbours outside the imaged window. By removing the triangulation at the border, we thus improve the calculation of the spatial and angular correlation functions.

\subsection{III.2. Translational and orientational correlation functions}
Translational and orientational correlation functions [\textit{S7}], $G_{K}(r)$ and $G_{6}(r)$, are defined from the translational and orientational order parameters, $\Psi_{K}(r)$ and $\Psi_{6}(r)$, as [\textit{S8}],
\begin{eqnarray*}
  G_{K} (r) &=& < \Psi_{K}(r) \Psi^{*}_{K}(0)> = \dfrac{1}{6} \sum_{l}^{6}\dfrac{1}{N(r)} \sum_{i,j}^{N(r)}\Psi_{K_{l}}(r_{i})\Psi^{*}_{K_{l}}(r_{j}) ,  \hspace{0.5cm} \Psi_{K_{l}}(r_{i}) = e^{i\mathbf{K_{l}r_{i}}} \\
  G_{6} (r) &=& < \Psi_{6}(r) \Psi^{*}_{6}(0)> = \dfrac{1}{N(r)} \sum_{i,j}^{N(r)} \Psi_{6}(r_{i}) \Psi^{*}_{6}(r_{j}),  \hspace{2cm} \Psi_{6}(r_{i}) = \dfrac{1}{N_{N}^{i}}\sum_{k}^{N_{N}^{i}} e^{i6\theta(r_{ik})}
\end{eqnarray*} 
where r is the distance of any lattice site to the origin, $N(r)$ is the number of vortex pairs separated by a distance r, $N_{N}^{i}$ is the number of the nearest neighbours of the vortex $i$ as given by the Delaunay triangulation, $\textbf{K}_{\textbf{l}}$ stands for each of the six main reciprocal lattice vectors and $\theta(r_{ik})$ is the angle of the nearest-neighbours bond between vortices i and j with respect to the reference axis. $G_{K}(r)$ and $G_{6}(r)$ functions shown in the publication (Fig. 3) are calculated using the above expressions from the vortex positions. The six main reciprocal lattice vectors $\mathbf{K}$ are determined from the position of the Bragg peaks in the Fourier transforms of the vortex position maps (see Section \textit{IV}). 

\hspace*{0.4cm}

We confirm that the calculated $G_{K}(r)$ and $G_{6}(r)$ for a perfect hexagonal lattice are equal to 1 when r coincides with the distance to n-th nearest neighbours (Fig. \textit{S2}). The relevant parameter for the discussion in the publication is the upper envelope of $G_{K} (r)$ and $G_{6} (r)$. 

\begin{figure}[h!]
\includegraphics[width=13cm,clip]{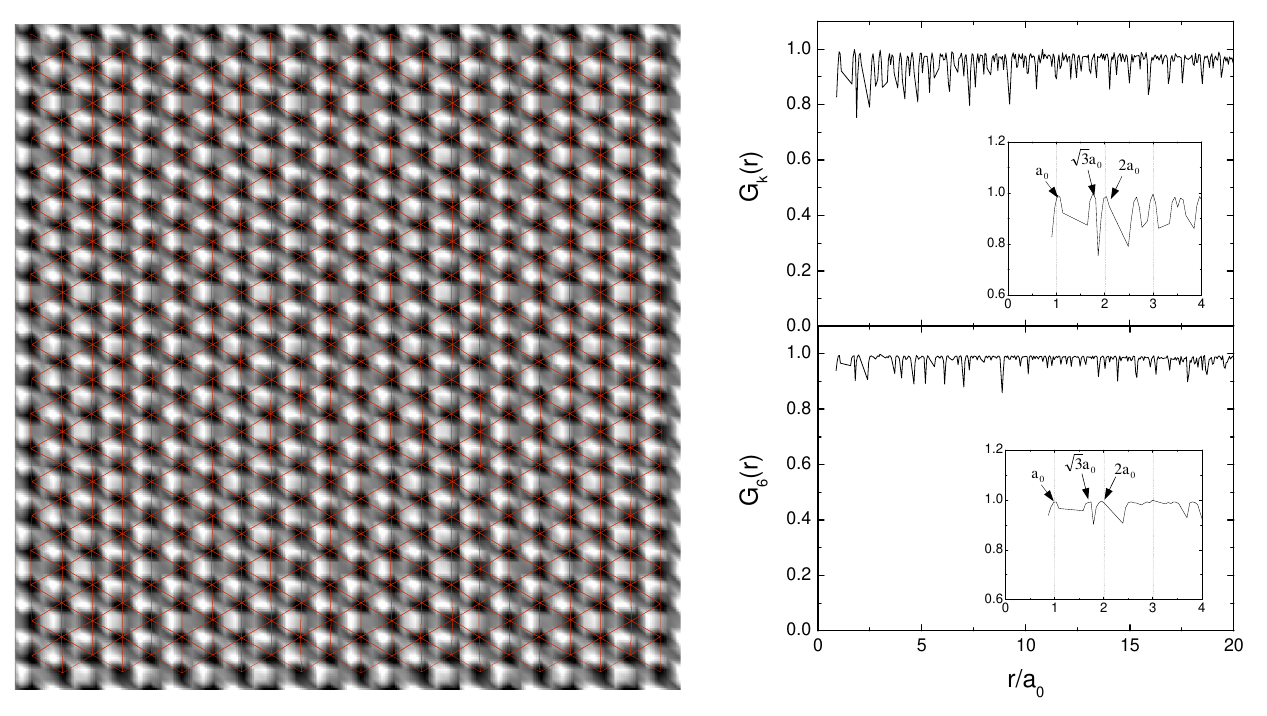}
\caption{\textbf{$G_{K}(r)$ and $G_{6}(r)$ functions for a perfect hexagonal lattice.} Left panel shows a perfect simulated hexagonal vortex lattice and right panel the calculated translational (upper) and orientational (bottom) correlation functions. Insets in right panels show a zooms-up of both correlation functions at distances up to $4 a_{0}$. The three first maxima appear at $r=a_0$, $r=\sqrt{3}a_0$ and $r\approx 2a_0$ as  it is expected for a lattice with long range order.}
\end{figure}

The position and sharpness of the peaks in the correlation functions are related to the averaged location of the nearest neighbours. For example, in Fig.3a of the publication, the first three peaks in correlation functions are located at $r=a_0$, $r=\sqrt{3}a_0$  and $2a_{0}$ at 0.5 T (insets in Fig. \textit{S2}). At 5.5 T, the second and third peaks merge in a single peak showing that order has become short range.

\hspace*{0.4cm}

To obtain $G_{K}(r)$ given in the publication, we average $G_{K_{l}}(r)$ over main reciprocal lattice directions. Fig. \textit{S3} shows $G_{K_{l}}(r)$ at 1.2 T and 3 T, which are in phase I (long range translational order) and phase II (short range translational order). The decay along different directions is very similar. Only in the ordered phase I we find slight variations (Fig. \textit{S3}a) that are not relevant for the discussion.  

\begin{figure}
\includegraphics[width=13cm,clip]{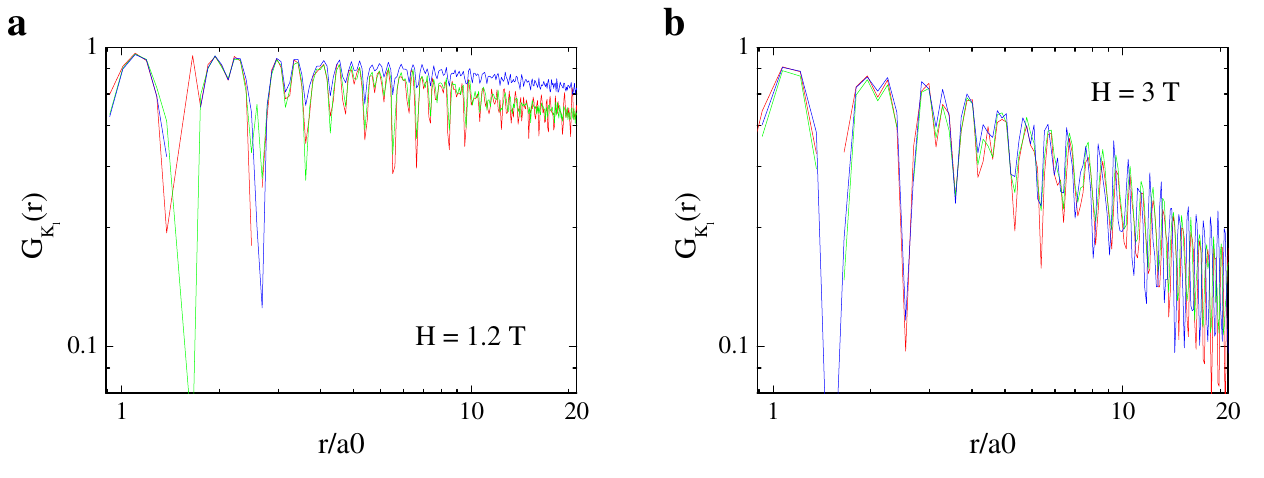}
\caption{\textbf{$G_{K_{l}}(r)$ functions in the translationally ordered and disordered phases.} $G_{K_{l}}(r)$ are calculated over the main reciprocal lattice directions $K_{1}$ (red), $K_{2}$ (green) and $K_{3}$ (blue) and averaged to give $G_{K}(r)$ functions shown in the publication. \textbf{a} and  \textbf{b} show, respectively, $G_{K_{l}}(r)$ at 1.2 T and 3 T where translational order goes from long range to short range.}
\end{figure}

\subsection{III.3.Vortex density maps}

To calculate the vortex density $\rho(x,y)$, we define a squared area centered at $(x,y)$ with a lateral size of 3 times the lattice parameter $a_{0}$. We count the number of vortices in this area, and divide by $A=(3a_{0})^{2}$. For the points $(x,y)$ at a distance to the edges below $3a_{0}/2$, the local vortex density is calculated over a smaller area. These points are not taken into account to calculate the histograms of the vortex density shown in Fig. 2c of the manuscript. In order to decrease the discontinuity between nearby pixels in $\rho(x,y)$ maps produced by the discrete nature of the vortex arrangement, they are smoothed using a Gaussian low-pass filter, which is the same for all the fields.

 \begin{figure}[h!]
\includegraphics[width=9.5cm,clip]{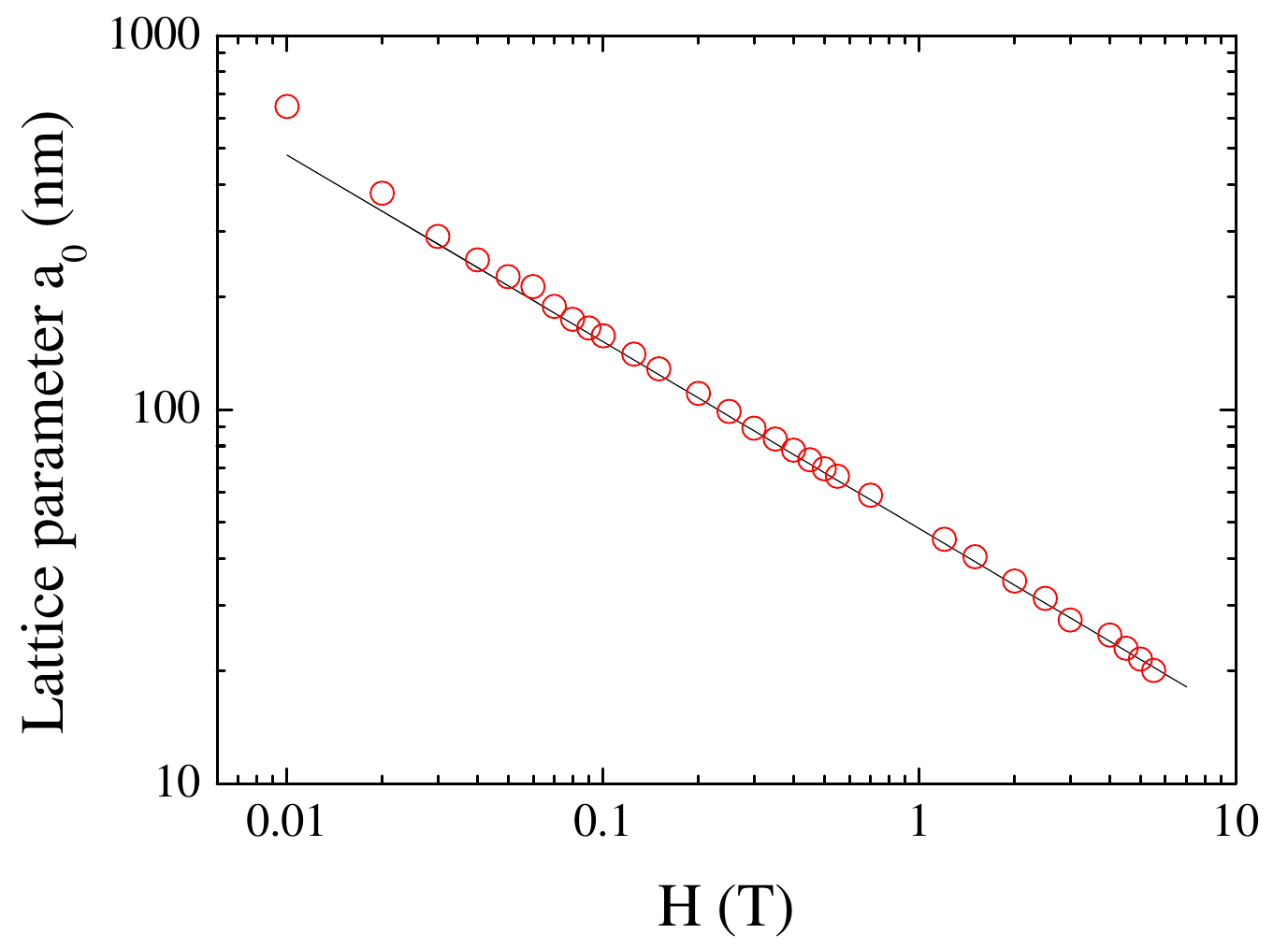}
\caption{\textbf{Intervortex distance vs magnetic field.} Red points are the field dependent lattice parameter $a_{0}$ determined as the position of the first maximum in the histograms of the intervortex distance which are obtained from the vortex positions at each value of the magnetic field. Black line is the expected intervortex distance for a superconductor with a hexagonal vortex lattice $a_{0} = (1.25)^{1/4}(\phi_{0}/H )^{1/2}$, where $\phi_{0}$ is the flux quantum.}
\end{figure}

The $\rho(x,y)$ maps shown in the publication are normalized by the averaged vortex density which is obtained from the total number of vortices in the image divided by the total area. Note that the mean vortex density is in good agreement with the expected value for a hexagonal lattice at all magnetic fields. Fig. \textit{S4} shows the field dependence of $a_{0}$ which is proportional to the mean vortex density. $a_{0}$  is experimentally determined as the position of the first maximum in the histograms of the intervortex distances obtained from the vortex positions at each field.

\subsection{III.4. Relative displacement correlator B(r)}
Deviations of the vortex arrangement with respect to a perfect hexagonal lattice are quantified by the relative displacement correlator $B(r)$ defined as $B(r)=\langle [\textbf{u}(\textbf{r})-\textbf{u}(\textbf{0})]^{2}\rangle/2$. $\textbf{u}(\textbf{r})= \textbf{r} - \textbf{r}_{\textbf{p}}$ is the displacement of each vortex at $\textbf{r}$ relative to its position in the perfect lattice $\textbf{r}_{\textbf{p}}$. In order to calculate $B(r)$, a perfect hexagonal lattice was generated using the reciprocal lattice vectors obtained from the Fourier transform of the vortex position maps (see Section \textit{IV}). The perfect hexagonal lattice is then compared to the real vortex arrangements in regions where all the vortices have 6 first neighbours. This allows to establish an unequivocal correspondence between each vortex and one single site in the perfect lattice. Finally, the average over the mean squared displacements between the real and perfect lattices, $\langle \vert \textbf{u}(\textbf{r})\vert^{2}\rangle$, is minimized respect to translation and rotation of the perfect lattice to obtain the best match between the two arrangements [\textit{S9,S10}]. The positions for the perfect lattice obtained after minimizations are used to compute the histograms and maps of the relative displacement, $u(r)$, and correlation functions $B(r)$ discussed in section \textit{VI}. To avoid errors related to discretization, we again smooth the relative displacement maps using a Gaussian function, as in Section \textit{III}.3.

\subsection{III.5. Mean squared variations of disorder potential}

Mean squared variation of disorder 1D potential, $\langle[V_{1D}(r)-V_{1D}(r')]^{2}\rangle$, are calculated by taking $V_{1D}(r)=z(r)\cdot\varepsilon_{L}$ where $z(r)$ is the topography measured simultaneously to vortex lattice maps and $\varepsilon_{L}= (\phi_{0}^{2}/4\pi\mu_{0}\lambda^{2})ln(\lambda/\xi)$ is the vortex energy per unit length. 

\hspace*{0.4cm}

Values for the coherence length and London penetration depth in the W nanodeposits are, respectively, $\xi = 6.25 nm$ and $\lambda = 850 nm$. The average in $\langle[V_{1D}(r)-V_{1D}(r')]^{2}\rangle$ is made over all vortex positions $r, r'$. Fig. \textit{S5}a shows $V_{1D}(r)$ map and vortex lattice positions taken simultaneously at 1.2 T. The resulting mean squared variation of $V_{1D}(r)$ is shown in Fig. 3d of the publication.

\hspace*{0.4cm}

To fit and extrapolate the data at larger distances, we simulate topography and $V_{1D}(r)$ maps using a periodic function in a region 100 times larger than the experimental area studied by STM, i.e, $10  \times 10 \mu m^{2}$ (Fig. \textit{S5}b). We also generate a hexagonal lattice over the same region using the reciprocal lattice vectors obtained from the Fourier transform of vortex position maps (see in Section \textit{IV}). A zoom up of the simulated $V_{1D}(r)$ and vortex lattice maps over a region with similar size than the experimental topography is shown in Fig. \textit{S5}b. We calculate the mean squared variation of simulated disorder potential at the generated vortex lattice positions to compare it with experiment (red line in Fig. 3d of the publication). 

\begin{figure}[h!]
\includegraphics[width=14cm,clip]{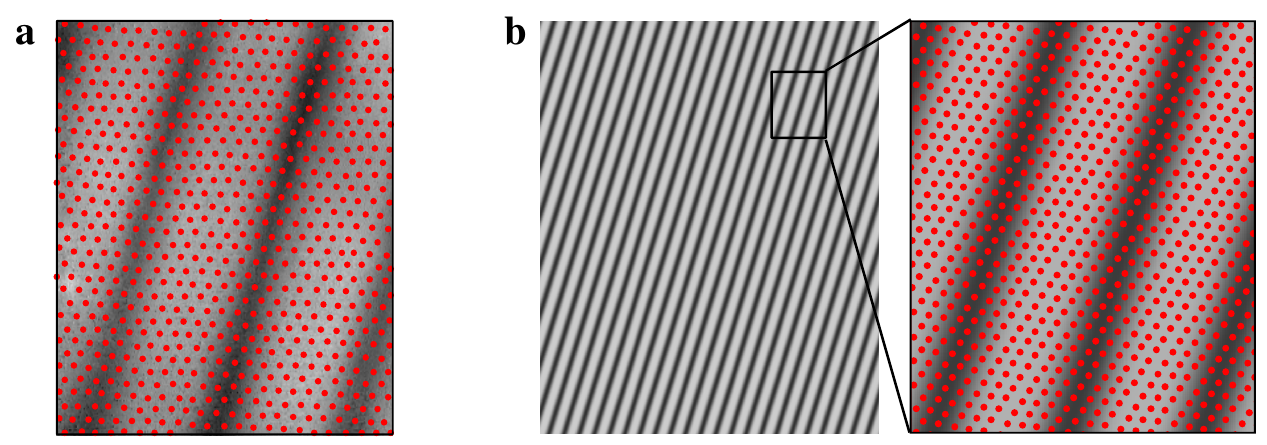}
\caption{\textbf{Correlations in vortex lattice disorder potential.} \textbf{a} shows a $V_{1D}(r)$ map obtained from spatial changes of vortex energy due to 1D modulation in the film thickness. Red points mark vortex positions obtained from the conductance maps measured at 1.2 T simultaneously to the topography. \textbf{b} Simulated $V_{1D}(r)$ in a region of $10  \times 10 \mu m^{2}$. A hexagonal vortex lattice  covering the same area has been generated using the experimental reciprocal lattice vectors. For clarity, this is only shown over a much smaller area of about the same size as the region studied by STM (\textbf{a}).}
\end{figure}

\section{IV. Order-disorder transition in the reciprocal space}

We have studied the order-disorder transition in the reciprocal space. Structural transitions of the vortex lattice in the Fourier space have been previously addressed through neutron diffraction experiments [\textit{S11}]. Bragg peaks obtained by neutron scattering mix structural information with vortex size or fluctuations. By contrast, our direct images and their triangulation provide the exact positions of vortices. Thus, the Fourier transforms (FT) of images consisting of solely the vortex position only contains information about the structural order of the vortex arrangement. 

\hspace*{0.4cm}

To extract the information about the modifications in the vortex arrangements, we make binary $512 \times 512$ maps from the vortex positions found in the conductance maps and study their FT. To compare the field induced changes in FT, binary maps at all magnetic fields contain the same number of vortices. Each vortex makes up one single pixel in the binary maps so that the form factor of a single vortex is a delta function (Fig. \textit{S6}a). 

\hspace*{0.4cm}

Fig. \textit{S6}a shows the FT of the binary map for the vortex positions at 1.2 T. In order to measure the magnetic field dependence of the sharpness of the Bragg peaks in the radial direction, we make circular profiles along the radius at which we observe the six main Bragg peaks. An example is shown the inset of Fig. \textit{S6}b for the FT at 1.2 T. The angular average of the six Bragg peaks is shown in the figure along with the fit to a Lorentzian distribution. 

\begin{figure}[h!]
\includegraphics[width=10cm,clip]{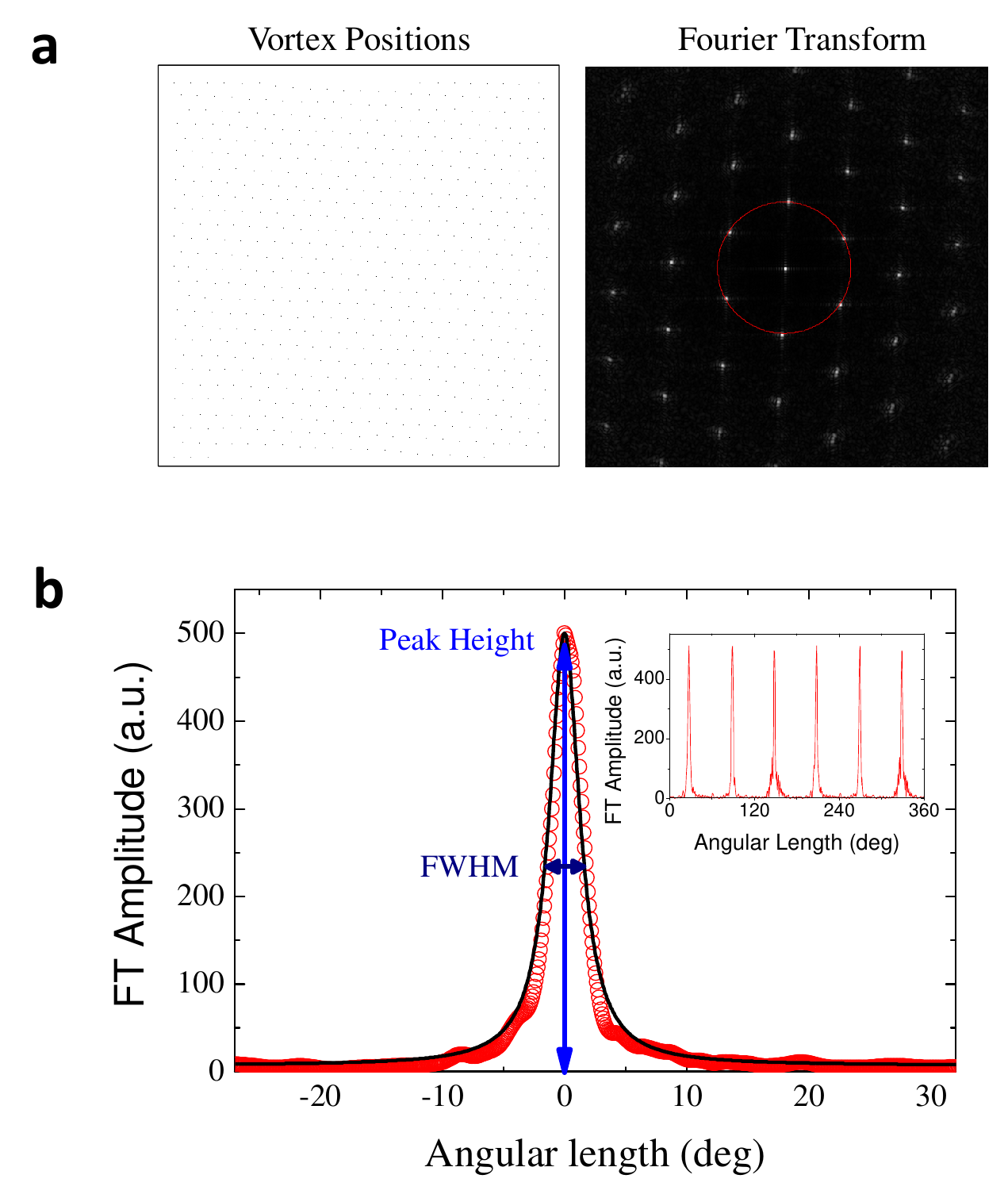}
\caption{\textbf{Fourier Transforms and Bragg peak profiles.} Left panel in \textbf{a} shows a $512 \times 512$ map of the vortex positions (black pixels) found at 1.2 T in a region containing around 600 vortices. Right panel in \textbf{a} shows the FT of the map in left panel. Red circle is the circular profile shown in the inset of \textbf{b} which shows the Fourier amplitude of the six main Bragg peaks along the circular path. The average of the six Bragg peaks is shown in \textbf{b} (red points). The black line is the fit to a Lorentzian distribution. Light and dark blue arrows mark, respectively, the maximum height of the peak and the FWHM obtained from the fit.}
\end{figure}

\hspace*{0.4cm}

Fig. \textit{S7}a shows FT of binary binary maps corresponding to the vortex images in Fig. 2a of the publication. The magnetic field dependence of the height and full width at half maximum (FWHM) of the averaged Bragg peaks as a function of the magnetic field is shown in Fig. \textit{S7}b. They give different behaviour in reciprocal space for each phase discussed in the publication. 

\hspace*{0.4cm}

In phase I, the FT shows six sharp Bragg peaks and higher harmonics. The amplitude of the first harmonic Bragg peaks slowly decreases with the magnetic field. But the FWHM is field independent (inset of Fig. \textit{S7}b). 

\hspace*{0.4cm}

In phase II, the hexagonal symmetry remains, but first and higher order harmonics lose intensity and become gradually more arc like instead of point like. The Fourier peak height drops faster with the magnetic field above 2 T, and the FWHM starts to increase slowly up to 4 T. 

\hspace*{0.4cm}

Then, in phase III, the peaks become arc-like nearly extending over a circle. The Fourier amplitude is strongly suppressed and the FWHM diverges. 

\begin{figure}
\includegraphics[width=16cm,clip]{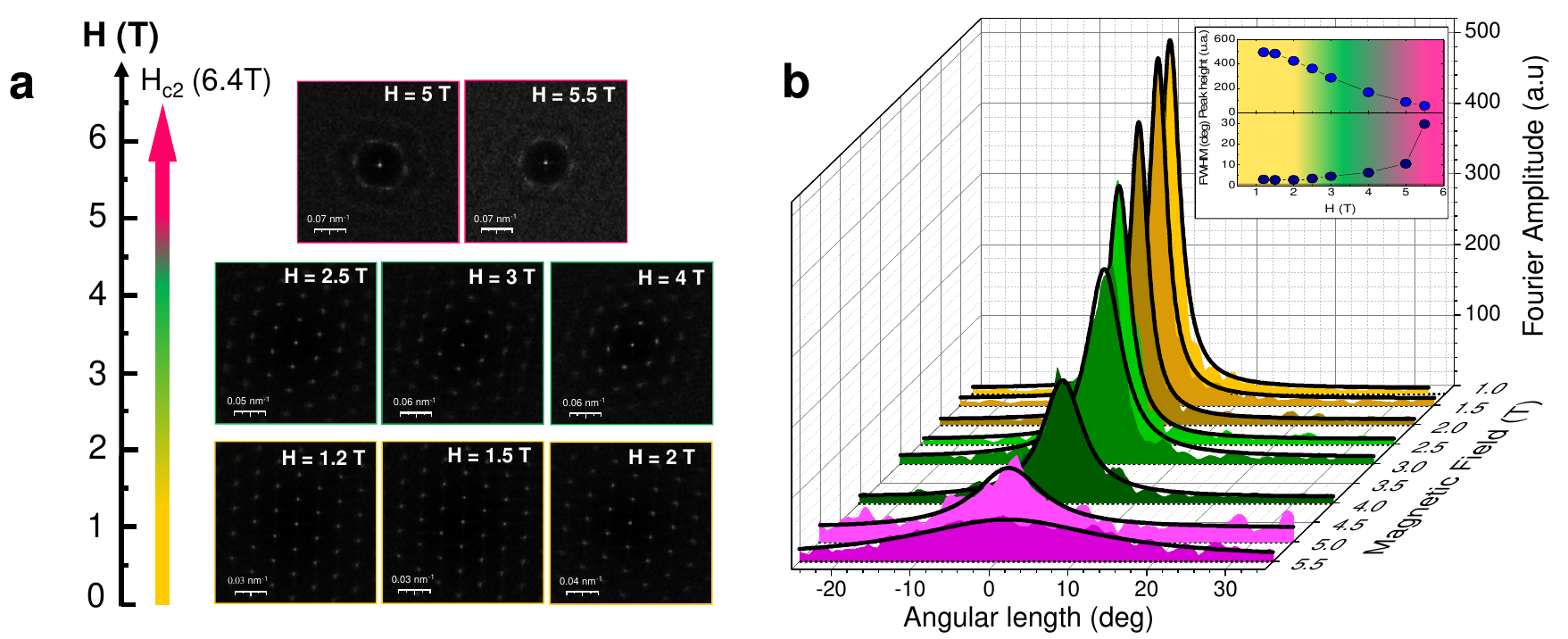}
\caption{\textbf{Order-disorder transition in reciprocal space.} (\textbf{a}) FT of vortex positions corresponding to STS images shown in Fig. 2a of the publication. In phase I (yellow framed FT), where the vortices are surrounded by six first neighbours, FT images show sharp Bragg peaks with hexagonal symmetry. In phase II (green framed FT),  FT images show a continuous broadening of the Bragg peaks into arcs as a consequence of the of appearance of dislocations and dislocations pairs. In phase III (in magenta), FT images show a circular structure reflecting the lost of orientacional order due to presence of large regions of bunched five-fold and seven-fold coordinated vortices. (\textbf{b}) shows the magnetic field evolution of the circular average of the six main Bragg peaks. Continuous lines are Lorentzian distribution fits to the averaged Bragg peaks. The peak height and FWHM are shown in the inset as a function of the magnetic field.}
\end{figure}

\section{V. Magnetic field dependence of the vortex interaction}

Vortex interaction in superconducting films with thickness much smaller than the London penetration depth, $\lambda$, the so-called Pearl interaction [\textit{S12}], is mainly dominated by the magnetic stray field. In this limit, the relevant length for the screening of the in-plane supercurrents is given by the effective 2D penetration depth $\Lambda = \lambda^{2}/d$.  An analytical expression for the Pearl interaction potential is [\textit{S13}],
\begin{equation*}
U_{v}(r) = \dfrac{\phi_{0}^{2}}{4\Lambda \mu_{0}}\left[\textbf{H}_{0}\left(\dfrac{r}{2\Lambda}\right)-\textbf{Y}_{0}\left(\dfrac{r}{2\Lambda}\right)\right]
\end{equation*}
where $\textbf{H}_{0}$ and $\textbf{Y}_{0}$ are, respectively, the Struve function and Bessel function of the second kind of order zero.   
 
\hspace*{0.4cm}
  
In the W-based nanodeposits, the London penetration depth ($\lambda = 850$ $nm$) is more than four times higher than the film thickness ($d = 200$ $nm$), and thus vortex interaction is well described by the above expression. Fig.\textit{S8} shows the magnetic field dependence of vortex interaction potential at a distance equal to $a_{0}$. The variation with the magnetic field comes from the field dependence of $a_{0}$, $\lambda$ and $\xi$ which are described within Ginzburg-Landau theory as $a_{0}=(1.25)^{1/4}(\phi_{0}/H )^{1/2}$, $\lambda_{H}=\lambda_{H=0}/(1-(H/H_{c2})^{2})^{1/2}$ and $\xi_{H}=\xi_{H=0}/(1-(H/H_{c2})^{2})^{1/2}$, respectively. In our W-based superconducting thin film, vortex interaction shows a maximum around 1.5 T. Then, it continuously decreases when magnetic field approaches to $H_{c2}$ ($= 6.4$ $T$) showing a stronger field dependence above 4 T. Note that first topological defects in our vortex images appear at magnetic fields (2 T) when U$_{v}$ decreases. 
  
\begin{figure}
\includegraphics[width=12cm,clip]{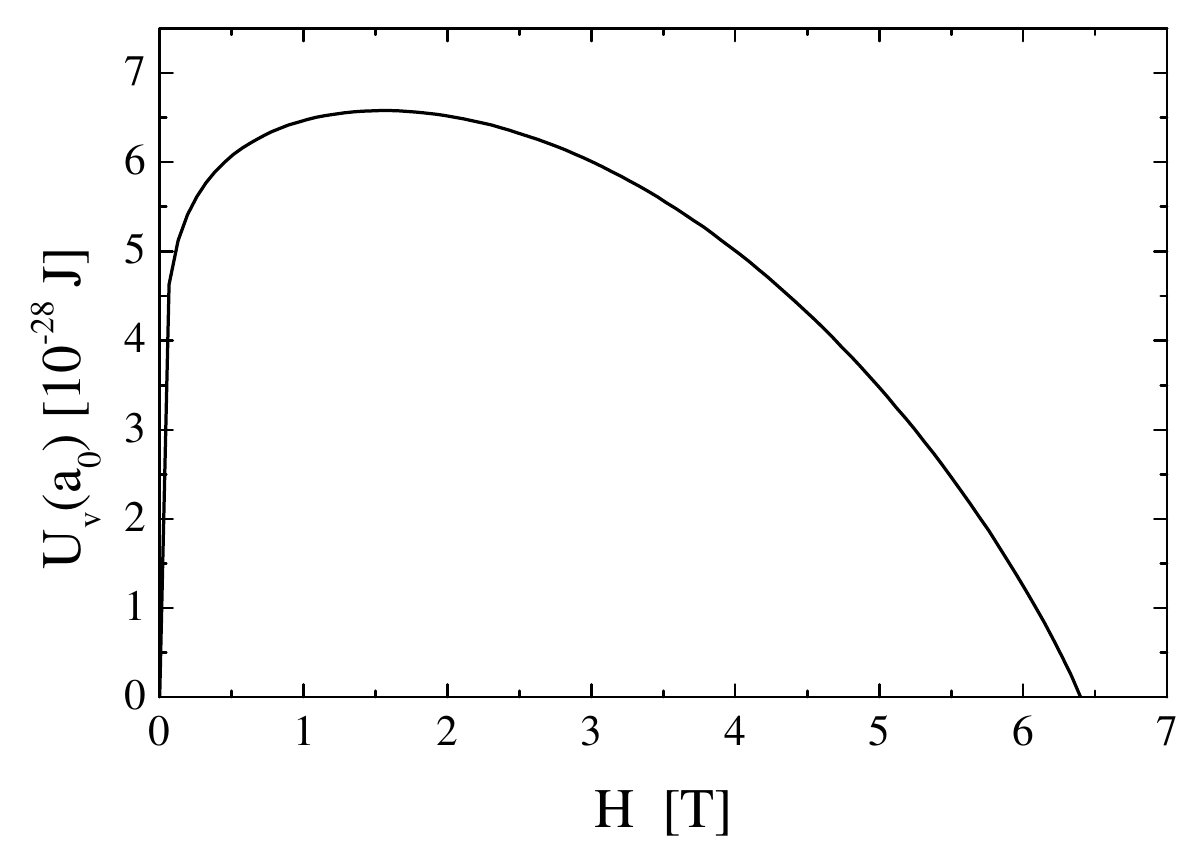}
\caption{\textbf{Intervortex interaction vs magnetic field.} Field dependence of the interaction potential between two vortices, $U_{v}(r)$,  at $r = a_{0}$ in the W superconducting thin film studied here.}
\end{figure}

\section{VI. $B(r)$ correlation functions and $u(r)$ maps for the 2D lattice in W thin films}

Fig. \textit{S9} shows the relative displacement $u(r)$ maps and histograms and $B(r)$ correlation functions for the vortex lattice images between 0.5 T and 2 T (Fig. 2a in the publication). Spatial maps of $u(r)$ (Fig. \textit{S9}a) show deviations of vortex positions relative to those in a perfect hexagonal lattice that appear randomly distributed over the surface. Histograms of $u(r)$ (Fig. \textit{S9}b) show that the amplitude of the vortex displacements roughly follows a Gaussian distribution with the mean value and width increasing with the magnetic field from 0.5 T to 2 T.

\begin{figure}
\includegraphics[width=16cm,clip]{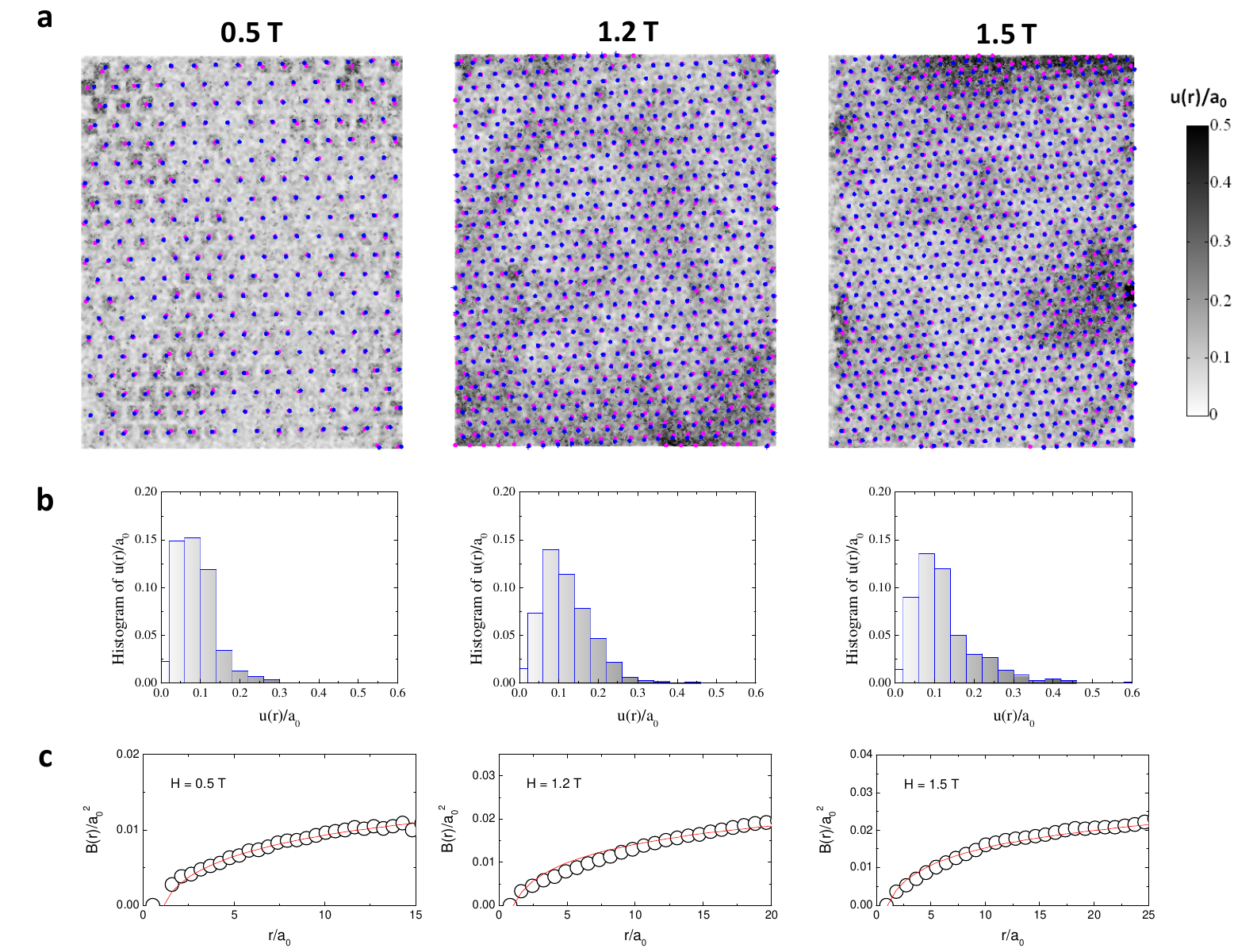}
\caption{\textbf{Spatial maps, histograms and correlation functions $B(r)$ of the relative displacement $u(r)$.} The figure shows relative displacement maps (\textbf{a}), histograms (\textbf{b}) and correlation functions $B(r)$ (\textbf{c}) for the vortex lattice images at 0.5 T, 1.2 T and 1.5 T. In \textbf{a}, real vortex positions and coordinates of  the perfect hexagonal lattice that minimized the mean squared deviation are marked, respectively, as magenta and blue dots. The color contrast in the images reflects the spatial variations of $u(r)$ according to the color scale shown on the right. The same color code is used in the histograms of u(r) shown in \textbf{b}. \textbf{c} shows the relative displacement correlation functions $B(r)$ calculated as described in section \textit{III.4}. Red  lines are the fits to logarithmic behaviour.} 
\end{figure}
 
\hspace*{0.4cm}

Relative displacements correlation functions $B(r)$ obtained from the vortex images and simulated perfect lattices as described in section \textit{III}.4 are shown in Fig. \textit{S9}c (black circles). We find that, in phase I, $B(r)$ grows as $ln(r/a_{0})$ and that is related to the translational correlation function through a Gaussian approximation ($G_{K}(r) \approx e^{-K^{2}B(r)/2}$) as described in the manuscript [\textit{S10}, \textit{S14} and Ref. 26 in the publication]. In phase I, $G_{K}(r) = (r/a_{0})^{-\eta_{K}}$ so that Gaussian relation gives (red lines in Fig. \textit{S9}c),  
\begin{equation*}
B(r)/a_{0}^{2} \approx \dfrac{3}{8\pi^{2}}\eta_{K} ln(r/a_{0}) 
\end{equation*} 

In the Fig. \textit{S9}c and in the Fig. 3c of the publication, we show B(r) functions as obtained directly form the vortex lattice images (black points) with those ones obtained using the above expression and the exponents $\eta_{K}$ (red lines). As we discuss in the manuscript, the agreement is very good. 

\hspace*{0.4cm}

Note that in presence of correlated pinning disorder, such as the disorder caused by T$_c$ variations, voids, crystal lattice dislocations or grain boundaries, $B(r)$ in 2D systems grows as $ln^{2}(r/a_{0})$ [\textit{S15} and Refs. 6,19 in the publication]. As we discuss previously, our sample is fully amorphous, with a homogeneous distribution of the composition and a homogeneous critical temperature. There are no voids, grain boundaries, dislocations or other distortions viewed in our images nor in TEM studies. Also, there is no visible effect on the vortex lattice other than the 1D modulation. Thus, the disorder in our sample is uncorrelated.

\hspace*{0.4cm}

This, in turn means that vortex displacements $B(r)$ have a weaker dependence on the distance ($ln(r/a_{0}$) instead of $ln^{2}(r/a_{0})$) than in the case of correlated disorder. In the case of correlated disorder, the gain of disorder energy is always larger than the attraction energy of dislocations pairs. An immediate consequence of this is that the 2D lattice is unstable to the formation of free dislocations and no true order-disorder transition can exist. This is the case of $Mo_{x}Ge$, $Mo_{x}Si$ and similar thin films where collective pinning leads to a phenomenology often characterized in terms of peak effect and irreversible behaviour which depends on the sample growth methodology [\textit{S5},\textit{S16},\textit{S17}]. In our amorphous and homogeneous thin film, uncorrelated disorder generates weak vortex displacements, and correspondingly the response of the lattice is reversible, without a peak effect nor irreversible behaviour found in four probe electrical transport and critical current measurements. The fundamental property is however the logarithmic distance dependence of $B(r)$. This, together with the vortex density fluctuations and the Gaussian relation between the translational correlation function $G_{K}(r)$ and $B(r)$, give three independent evidences that the disorder potential is uncorrelated and random.

\section{VII. Estimation BKTHNY 2D melting transition in our W-based thin film at 0.1 K}

We have calculated the thermal melting line predicted within BKTHNY theory. The BKTHNY transition temperature, above which the 2D solid becomes unstable to the proliferation of dislocation, is given by the condition [\textit{S18}], 
\begin{equation*}
\dfrac{A\mu a_{0}^{2}d}{K_{B}T} = 4 \pi
\end{equation*}
where $\mu$ is the shear modulus of the lattice and $A = 0.65$ is a constant that accounts for the renormalization of $\mu$ at the transition. The shear modulus depends on both temperature and magnetic field as 
\begin{equation*}
\mu=\dfrac{H_{c}(T)^{2}}{4\mu_{0}} b(1-0.58b+0.29b^{2})(1-b)^{2}
\end{equation*}
where $H_{c}(T)=H_{c2}(T)/\sqrt{2}\kappa(1.25-0.25t)^{2}$, with $b=H/H_{c2}(T)$, $t=T/T_{c}$ and $\kappa$ the Ginzburg-Landau parameter [\textit{S18}]. These expressions give the melting transition line directly from the properties the  superconducting thin film without any fitting parameter. Using the values for our W-based thin film given in Section \textit{III}.5, we find that the thermal melting transition line crosses 0.1 K at 6.2 T, which is far above the fields at which we observe the order-disorder transition.   

\section{VIII. Estimation quantum melting transition in our W-based thin film at 0.1 K}

The possible appearance of quantum fluctuations in superconducting thin films with $d > \xi$ can be discussed through the magnitude of the resistance ratio $\rho_{n}/dR_{q}$, where $R_{q}=\hbar/e^{2} = 4.1 k\Omega$ is the quantum resistance and $\rho_{n}$ is the normal state resistivity [\textit{S19}]. The quantum melting line is given by
\begin{equation*}
B_{m}^{q} = B_{c2} \left[1-\dfrac{2}{\pi}exp\left(\dfrac{2\pi}{3}\alpha - \dfrac{\alpha^{2}}{2} - \dfrac{\pi^{3}c_{L}^{2}}{2}\dfrac{R_{q}}{R_{n}^{*}}\right)\right],
\end{equation*}
where $c_{L}$ is the Lindemann number, $\alpha = 2/\sqrt{\pi}\nu$, $\nu$ a numerical constant of order unity and $R_{n}^{*} = \rho_{n}/a_{0}$ the effective sheet resistance.  Using $\rho_{n} = 275$ $\mu \Omega cm$ for our W-based thin film, and taking $c_{L} = 0.1-0.3$ [\textit{S19}], we find that the quantum melting transition line at 0.1 K is above 6.1 T, which is again much higher than the fields at which we observe the order-disorder transition. This value is of the same order than the fields at which quantum melting has been observed through transport measurements in other superconducting thin films with similar parameters(\textit{S20}, \textit{S21}).

\section{IX. Brief description of supplementary videos}

\paragraph{Supplementary Video S1: LatticeCompression.avi.}
Whole sequence of the vortex images shown in Fig. 1 and Fig. 2 of the publication when increasing the magnetic field from 0.01 T to 5.5 T at 0.1 K.

\hspace*{0.5cm}

\paragraph{Supplementary Video S2: OrderDisorderTransition.avi.}
Order disorder transition in the 2D vortex lattice when increasing the magnetic field from 0.5 T to 5.5 T at 0.1 K. The video presents the vortex lattice images and Fourier transforms shown in Fig. \textit{S7} along with histograms showing the number of 5-fold, 6-fold and 7-fold coordinated vortices.

\hspace*{3cm}

\section{Supplementary references}

\renewcommand{\labelenumi}{S\arabic{enumi}}

\begin{enumerate}

\item Suderow H., Guillam\'on I. \& Vieira S. Compact very low temperature scanning tunneling microscope with mechanically driven horizontal linear positioning stage. \emph{Review of Scientific Instruments} \textbf{82}, 033711 (2011).

\item Guillam\'on I., Suderow H., Vieira S., Fern\'andez-Pacheco A.,  Ses\'e J., C\'ordoba R., De Teresa J.M.\& Ibarra M.R. Nanoscale superconducting properties of amorphous W-based deposits grown with focused-ion-beam. \emph{New Journal of Physics} \textbf{10}, 093005 (2008).

\item Guillam\'on I., Suderow H., Fern\'andez-Pacheco A.,  Ses\'e J., C\'ordoba R., De Teresa J.M., Ibarra M.R \& Vieira S. Direct observation of melting in a two-dimensional superconducting vortex lattice. \emph{Nature Physics} \textbf{5}, 651-655 (2009).

\item Guillam\'on I., Suderow H., Vieira S., Cario L., Diener P. \& Rodi\'ere P. Superconducting density of states and vortex cores of 2H-NbS$_{2}$. \emph{Physical Review Letters} \textbf{101}, 166407 (2008).

\item Kes P. \& Tsuei C. Two-dimensional collective flux pinning, defects, and structural relaxation in amorphous superconducting films. \emph{Physical Review B} \textbf{28}, 5126 (1983).

\item Yazdani A. \emph{et al.} Observation of Kosterlitz-Thouless-type melting of the disordered vortex lattice in thin films of a-MoGe. \emph{Physical Review Letters} \textbf{70}, 505 (1993).

\item Nelson D.R, Rubinstein M. \& Spaepen F. Order in two-dimensional binary random arrays. \emph{Philosophical Magazine A} \textbf{49}, 1982 (1982).

\item Nelson D.R \& Halperin B.I. Dislocation-mediated melting in two dimensions. \emph{Physical Review B} \textbf{19}, 2457 (1979).

\item Marchevsky M., Keurentjes A., Arts J. \& Kes P.H. Elastic deformations in field-cooled vortex lattices in NbSe$_{2}$. \emph{Physical Review B} \textbf{57}, 6061 (1998).

\item Kim P., Yao Z., Bolle C.A \& Lieber M. Structure of flux lattices with weak disorder at large length scales. \emph{Physical Review B} \textbf{60}, R12 589 (1999).

\item Cubitt R., Forgan E.M. , Yang G., Lee S.L., Mck Paul D., Mook H.A., Yethiraj M., Kes P.H., Li T.W., Menovsky A.A., Tarnawski Z. \& Mortensen K. Direct observation of magnetic flux lattice melting and decomposition
in the high-T$_{c}$ superconductor $Bi_{2.15}Sr_{1.95}CaCu_{2}O_{8+x}$. \emph{Nature} \textbf{365}, 407 (1993).

\item Pearl J. Current distribution in superconducting films carrying quantized fluxoids. \emph{Applied Physics Letter} \textbf{5}, 65 (1964). 

\item Brandt E.H. Vortex-vortex interaction in thin superconducting films. \emph{Physical Review B} \textbf{79}, 134529 (2009).

\item Fasano Y. \& Menghini M. Magnetic-decoration imaging of structural transitions induced in vortex matter. \emph{Superconductor Science and Technology} \textbf{21}, 023001 (2008).

\item Nattermann T. \& Scheidl S.  Vortex-glass phases in type-II superconductors. \emph{Advanced Physics} \textbf{49}, 607-704 (2000).

\item W\"{o}rdenweber R., Kes P.H, \& Tsuei C.C.  Peak and history effects in two-dimensional collective flux pinning. \emph{Physical Review B} \textbf{33}, 3172 (1986).

\item W\"{o}rdenweber R., Pruymboom A., \& Kes P.H.  Characterization of defects in amorphous type-$II$ superconductors using the collective flux pinning properties. \emph{Journal of Low Temperature Physics} \textbf{70}, 253-278 (1988).

\item Geers J.M.E, Attanasio C., Hesselberth M.B.S., Aarts J. \& Kes P.H. Dynamic vortex ordering in thin $a-Nb_{70}Ge_{30}$ films. \emph{Physical Review B} \textbf{63}, 094511 (2001).

\item Blatter, G., Ivlev, B., Kagan, Y., Theunissen, M., Volokitin, Y. \& Kes, P.  Quantum liquid of vortices in superconductors at T=0. \emph{Physical Review B} \textbf{50}, 13013 (1994).

\item Okuma, S., Imamoto, Y. \& Morita, M.  Vortex Glass Transition and Quantum Vortex Liquid at Low Temperature in a Thick $a- Mo_{x}Si_{1-x}$ Film. \emph{Physical Review Letters} \textbf{86}, 3136-3139 (2001).

\item Kes, P.H., Theunissen, M.H. \& Becker, B.  Properties of the vortex liquid in the classical and quantum regime. \emph{Physica (Amsterdam)} \textbf{282C-287C}, 331 (1997).

\end{enumerate}
\end{widetext}

\begin{thebibliography}{10}
\expandafter\ifx\csname url\endcsname\relax
  \def\url#1{\texttt{#1}}\fi
\expandafter\ifx\csname urlprefix\endcsname\relax\def\urlprefix{URL }\fi
\providecommand{\bibinfo}[2]{#2}
\providecommand{\eprint}[2][]{\url{#2}}

\bibitem{Anderson58}
\bibinfo{author}{Anderson, P.~W.}
\newblock \bibinfo{title}{Absence of diffusion in certain random lattices}.
\newblock \emph{\bibinfo{journal}{Phys. Rev.}} \textbf{\bibinfo{volume}{109}},
  \bibinfo{pages}{1492} (\bibinfo{year}{1958}).

\bibitem{Fisher89}
\bibinfo{author}{Fisher, M.}, \bibinfo{author}{Weichman, P.},
  \bibinfo{author}{Grinstein, G.} \& \bibinfo{author}{Fisher, D.}
\newblock \bibinfo{title}{Boson localization and the superfluid-insulator
  transition}.
\newblock \emph{\bibinfo{journal}{Phys. Rev. B}} \textbf{\bibinfo{volume}{40}},
  \bibinfo{pages}{546} (\bibinfo{year}{1989}).

\bibitem{Wehr06}
\bibinfo{author}{Wehr, J.}, \bibinfo{author}{Niederberger, A.},
  \bibinfo{author}{Sanchez-Palencia, L.} \& \bibinfo{author}{Lewenstein, M.}
\newblock \bibinfo{title}{Disorder versus the {Mermin$-$Wagner$-$Hohenberg}
  effect: From classical spin systems to ultracold atomic gases}.
\newblock \emph{\bibinfo{journal}{Phys. Rev. B}} \textbf{\bibinfo{volume}{74}},
  \bibinfo{pages}{224448} (\bibinfo{year}{2006}).

\bibitem{Billy08}
\bibinfo{author}{Billy, J.} \emph{et~al.}
\newblock \bibinfo{title}{Direct observation of {Anderson} localization of
  matter waves in a controlled disorder}.
\newblock \emph{\bibinfo{journal}{Nature}} \textbf{\bibinfo{volume}{453}},
  \bibinfo{pages}{891--894} (\bibinfo{year}{2008}).

\bibitem{Sanchez-Palencia10}
\bibinfo{author}{Sanchez-Palencia, L.} \& \bibinfo{author}{Lewenstein, M.}
\newblock \bibinfo{title}{Disordered quantum gases under control}.
\newblock \emph{\bibinfo{journal}{Nat. Phys.}} \textbf{\bibinfo{volume}{6}},
  \bibinfo{pages}{87--95} (\bibinfo{year}{2010}).

\bibitem{Nattermann95}
\bibinfo{author}{Nattermann, T.}, \bibinfo{author}{Scheidl, S.},
  \bibinfo{author}{Korshunov, S.} \& \bibinfo{author}{Li, M.}
\newblock \bibinfo{title}{Absence of reentrance in the two-dimensional
  {XY}-model with random phase shift}.
\newblock \emph{\bibinfo{journal}{J. Phys. (France) I}}
  \textbf{\bibinfo{volume}{5}}, \bibinfo{pages}{565} (\bibinfo{year}{1995}).

\bibitem{Fertig95}
\bibinfo{author}{Cha, M.-C.} \& \bibinfo{author}{Fertig, H.~A.}
\newblock \bibinfo{title}{Disorder-induced phase transition in two-dimensional
  crystals}.
\newblock \emph{\bibinfo{journal}{Phys. Rev. Lett.}}
  \textbf{\bibinfo{volume}{74}}, \bibinfo{pages}{4867} (\bibinfo{year}{1995}).

\bibitem{HN79}
\bibinfo{author}{Nelson, D.~R.} \& \bibinfo{author}{Halperin, B.~I.}
\newblock \bibinfo{title}{Dislocation-mediated melting in two dimensions}.
\newblock \emph{\bibinfo{journal}{Physical Review B}}
  \textbf{\bibinfo{volume}{19}}, \bibinfo{pages}{2457} (\bibinfo{year}{1979}).

\bibitem{Berezinskii72}
\bibinfo{author}{Berezinskii, V.}
\newblock \bibinfo{title}{Destruction of long-range order in one-dimensional
  and two-dimensional systems possessing a continuous symmetry group. {II.
  Quantum Systems}}.
\newblock \emph{\bibinfo{journal}{Sov. Phys. JETP}}
  \textbf{\bibinfo{volume}{34}}, \bibinfo{pages}{610--616}
  (\bibinfo{year}{1972}).

\bibitem{kosterlitz73}
\bibinfo{author}{Kosterlitz, J.~M.} \& \bibinfo{author}{Thouless, D.~J.}
\newblock \bibinfo{title}{Ordering, metastability and phase transitions in
  two-dimensional systems}.
\newblock \emph{\bibinfo{journal}{J. Phys. C}} \textbf{\bibinfo{volume}{6}},
  \bibinfo{pages}{1181} (\bibinfo{year}{1973}).

\bibitem{Carpentier98}
\bibinfo{author}{Carpentier, D.} \& \bibinfo{author}{Doussal, P.~L.}
\newblock \bibinfo{title}{Melting of two-dimensional solids on disordered
  substrates}.
\newblock \emph{\bibinfo{journal}{Phys. Rev. Lett.}}
  \textbf{\bibinfo{volume}{81}}, \bibinfo{pages}{1881} (\bibinfo{year}{1998}).

\bibitem{Natterman90}
\bibinfo{author}{Nattermann, T.}
\newblock \bibinfo{title}{Scaling approach to pinning: charge density waves and
  giant flux creep in superconductors}.
\newblock \emph{\bibinfo{journal}{Phys. Rev. Lett.}}
  \textbf{\bibinfo{volume}{64}}, \bibinfo{pages}{2454} (\bibinfo{year}{1990}).

\bibitem{Minnhagen87}
\bibinfo{author}{Minnhagen, P.}
\newblock \bibinfo{title}{The two-dimensional {Coulomb} gas, vortex unbinding,
  and superfluid-superconducting films}.
\newblock \emph{\bibinfo{journal}{Rev. Mod. Phys.}}
  \textbf{\bibinfo{volume}{59}}, \bibinfo{pages}{1001--1066}
  (\bibinfo{year}{1987}).

\bibitem{Mermin68}
\bibinfo{author}{Mermin, N.~D.}
\newblock \bibinfo{title}{Crystalline order in two dimensions}.
\newblock \emph{\bibinfo{journal}{Phys. Rev.}} \textbf{\bibinfo{volume}{176}},
  \bibinfo{pages}{250} (\bibinfo{year}{1968}).

\bibitem{Hohenberg67b}
\bibinfo{author}{Hohenberg, P.}
\newblock \bibinfo{title}{Existence of long-range order in one and two
  dimensions}.
\newblock \emph{\bibinfo{journal}{Phys. Rev.}} \textbf{\bibinfo{volume}{158}},
  \bibinfo{pages}{383--366} (\bibinfo{year}{1967}).

\bibitem{HN78}
\bibinfo{author}{Halperin, B.~I.} \& \bibinfo{author}{Nelson, D.~R.}
\newblock \bibinfo{title}{Theory of two-dimensional melting}.
\newblock \emph{\bibinfo{journal}{Phys. Rev. Lett.}}
  \textbf{\bibinfo{volume}{41}}, \bibinfo{pages}{121} (\bibinfo{year}{1978}).

\bibitem{Young79}
\bibinfo{author}{Young, A.~P.}
\newblock \bibinfo{title}{Melting and the vector {Coulomb} gas in two
  dimensions}.
\newblock \emph{\bibinfo{journal}{Phys. Rev. B}} \textbf{\bibinfo{volume}{19}},
  \bibinfo{pages}{1855} (\bibinfo{year}{1979}).

\bibitem{Nelson83}
\bibinfo{author}{Nelson, D.~R.}
\newblock \bibinfo{title}{Reentrant melting in solid films with quenched random
  impurities}.
\newblock \emph{\bibinfo{journal}{Phys. Rev. B}} \textbf{\bibinfo{volume}{27}},
  \bibinfo{pages}{2902} (\bibinfo{year}{1983}).

\bibitem{Giamarchi00}
\bibinfo{author}{LeDoussal, P.} \& \bibinfo{author}{Giamarchi, T.}
\newblock \bibinfo{title}{Dislocation and {Bragg} glasses in two dimensions}.
\newblock \emph{\bibinfo{journal}{Physica C}} \textbf{\bibinfo{volume}{C331}},
  \bibinfo{pages}{233--240} (\bibinfo{year}{2000}).

\bibitem{Guillamon11a}
\bibinfo{author}{Guillamon, I.} \emph{et~al.}
\newblock \bibinfo{title}{Direct observation of stress accumulation and
  relaxation in small bundles of superconducting vortices in tungsten thin
  films}.
\newblock \emph{\bibinfo{journal}{Phys. Rev. Lett.}}
  \textbf{\bibinfo{volume}{106}}, \bibinfo{pages}{077001}
  (\bibinfo{year}{2011}).

\bibitem{Nattermann96}
\bibinfo{author}{Korshunov, S.} \& \bibinfo{author}{Nattermann, T.}
\newblock \bibinfo{title}{Phase diagram of a {Josephson} junction array with
  positional disorder}.
\newblock \emph{\bibinfo{journal}{Physica B}} \textbf{\bibinfo{volume}{222}},
  \bibinfo{pages}{280--286} (\bibinfo{year}{1996}).

\bibitem{SadrLahijany97}
\bibinfo{author}{SadrLahijany, M.}, \bibinfo{author}{Ray, P.} \&
  \bibinfo{author}{Stanley, H.}
\newblock \bibinfo{title}{Dispersity-driven melting transition in
  two-dimensional solids}.
\newblock \emph{\bibinfo{journal}{Phys. Rev. Lett.}}
  \textbf{\bibinfo{volume}{79}}, \bibinfo{pages}{3206} (\bibinfo{year}{1997}).

\bibitem{Abanin07}
\bibinfo{author}{Abanin, D.}, \bibinfo{author}{Lee, P.} \&
  \bibinfo{author}{Levitov, L.}
\newblock \bibinfo{title}{Randomness-induced {XY} ordering in a graphene
  quantum hall ferromagnet}.
\newblock \emph{\bibinfo{journal}{Phys. Rev. Lett.}}
  \textbf{\bibinfo{volume}{98}}, \bibinfo{pages}{156801}
  (\bibinfo{year}{2007}).

\bibitem{Niederberger08}
\bibinfo{author}{Niederberger, A.} \emph{et~al.}
\newblock \bibinfo{title}{Disorder-induced order in two-component
  {Bose-Einstein} condensates}.
\newblock \emph{\bibinfo{journal}{Phys. Rev. Lett.}}
  \textbf{\bibinfo{volume}{100}}, \bibinfo{pages}{030403}
  (\bibinfo{year}{2008}).

\bibitem{Radzihovsky01}
\bibinfo{author}{Radzihovsky, L.}, \bibinfo{author}{Frey, E.} \&
  \bibinfo{author}{Nelson, D.}
\newblock \bibinfo{title}{Novel phases and reentrant melting of two-dimensional
  colloidal crystals}.
\newblock \emph{\bibinfo{journal}{Phys. Rev. E}} \textbf{\bibinfo{volume}{63}},
  \bibinfo{pages}{031503} (\bibinfo{year}{2001}).

\bibitem{Giamarchi95}
\bibinfo{author}{Giamarchi, T.} \& \bibinfo{author}{LeDoussal, P.}
\newblock \bibinfo{title}{Elastic theory of flux lattices in the presence of
  weak disorder}.
\newblock \emph{\bibinfo{journal}{Phys. Rev. B}} \textbf{\bibinfo{volume}{52}},
  \bibinfo{pages}{1242} (\bibinfo{year}{1995}).

\bibitem{Zahn99}
\bibinfo{author}{Zahn, K.}, \bibinfo{author}{Lenke, R.} \&
  \bibinfo{author}{Maret, G.}
\newblock \bibinfo{title}{Two-stage melting of paramagnetic colloidal crystals
  in two dimensions}.
\newblock \emph{\bibinfo{journal}{Phys. Rev. Lett.}}
  \textbf{\bibinfo{volume}{82}}, \bibinfo{pages}{2721} (\bibinfo{year}{1999}).

\bibitem{Carpentier00}
\bibinfo{author}{Carpentier, D.} \& \bibinfo{author}{Doussal, P.~L.}
\newblock \bibinfo{title}{Topological transitions and freezing in {XY} models
  and {Coulomb} gases with quenched disorder: renormalization via traveling
  waves}.
\newblock \emph{\bibinfo{journal}{Nuc. Phys. B}}
  \textbf{\bibinfo{volume}{588}}, \bibinfo{pages}{565--629}
  (\bibinfo{year}{2000}).

\bibitem{Velarde09}
\bibinfo{author}{Herrera-Velarde, S.} \& \bibinfo{author}{von Gr\"{u}nberg,
  H.~H.}
\newblock \bibinfo{title}{Disorder-induced vs temperature-induced melting of
  two-dimensional colloidal crystals}.
\newblock \emph{\bibinfo{journal}{Soft Matter}} \textbf{\bibinfo{volume}{5}},
  \bibinfo{pages}{391--399} (\bibinfo{year}{2009}).

\end{thebibliography}
\end{document}